\documentclass[twocolumn]{aastex631}
\usepackage{color}
\usepackage[ruled, linesnumbered]{algorithm2e}
\usepackage{soul}
\usepackage{ulem}
\usepackage{tabularx}
\usepackage{makecell}
\usepackage{graphicx}

\graphicspath{{./}{figures/}}

\begin{document}

\title{Statistical Analyses of Solar Prominences and Active Region Features in 304 \AA \ Filtergrams detected via Deep Learning}

\correspondingauthor{Q. Hao}
\email{haoqi@nju.edu.cn}

\author[0009-0008-6298-2333]{T. Zhang}
\affiliation{School of Astronomy and Space Science, Nanjing
	University, Nanjing 210023, China}

\author[0000-0002-9264-6698]{Q. Hao}
\affiliation{School of Astronomy and Space Science, Nanjing
	University, Nanjing 210023, China}
\affiliation{Key Laboratory of Modern Astronomy and Astrophysics
	(Nanjing University), Ministry of Education, Nanjing 210023, China}

\author[0000-0002-7289-642X]{P. F. Chen}
\affiliation{School of Astronomy and Space Science, Nanjing
	University, Nanjing 210023, China}
\affiliation{Key Laboratory of Modern Astronomy and Astrophysics
	(Nanjing University), Ministry of Education, Nanjing 210023, China}
%% Note that the \and command from previous versions of AASTeX is now
%% depreciated in this version as it is no longer necessary. AASTeX
%% automatically takes care of all commas and "and"s between authors names.

%% AASTeX 6.31 has the new \collaboration and \nocollaboration commands to
%% provide the collaboration status of a group of authors. These commands
%% can be used either before or after the list of corresponding authors. The
%% argument for \collaboration is the collaboration identifier. Authors are
%% encouraged to surround collaboration identifiers with ()s. The
%% \nocollaboration command takes no argument and exists to indicate that
%% the nearby authors are not part of surrounding collaborations.

%% Mark off the abstract in the ``abstract'' environment.
\begin{abstract}

Solar active regions (ARs) are areas on the Sun with very strong magnetic fields where various activities take place. Prominences are one of the typical solar features in the solar atmosphere, whose eruptions often lead to solar flares and coronal mass ejections (CMEs). Therefore, studying their morphological features and their relationship with solar activity is useful in predicting eruptive events and in understanding the long-term evolution of solar activities. A huge amount of data have been collected from various ground-based telescopes and satellites. The massive data make human inspection difficult. For this purpose, we developed an automated detection method for prominences and ARs above the solar limb based on deep learning techniques. We applied it to process the 304 \AA \ data obtained by SDO/AIA  from 2010 May 13 to 2020 December 31. Besides the butterfly diagrams and latitudinal migrations of the prominences and ARs during solar cycle 24, the variations of their morphological features (such as the locations, areas, heights, and widths) with the calendar years and the latitude bands were analyzed. Most of these statistical results based on our new method are in agreement with previous studies, which also guarantees the validity of our method. The N-S asymmetry indices of the prominences and ARs show that the northern hemisphere dominates in solar cycle 24, except for 2012--2015,  and 2020 for ARs. The high-latitude prominences show much stronger N-S asymmetry that the northern hemisphere is dominant in $\sim$2011 and $\sim$2015 and the southern hemisphere is dominant during 2016--2019.
\end{abstract}

%% Keywords should appear after the \end{abstract} command.
%% The AAS Journals now uses Unified Astronomy Thesaurus concepts:
%% https://astrothesaurus.org
%% You will be asked to selected these concepts during the submission process
%% but this old "keyword" functionality is maintained in case authors want
%% to include these concepts in their preprints.
%\keywords{Classical Novae (251) --- Ultraviolet astronomy(1736) --- History of astronomy(1868) --- Interdisciplinary astronomy(804)}
%% From the front matter, we move on to the body of the paper.
%% Sections are demarcated by \section and \subsection, respectively.
%% Observe the use of the LaTeX \label
%% command after the \subsection to give a symbolic KEY to the
%% subsection for cross-referencing in a \ref command.
%% You can use LaTeX's \ref and \label commands to keep track of
%% cross-references to sections, equations, tables, and figures.
%% That way, if you change the order of any elements, LaTeX will
%% automatically renumber them.
%%
%% We recommend that authors also use the natbib \citep
%% and \citet commands to identify citations.  The citations are
%% tied to the reference list via symbolic KEYs. The KEY corresponds
%% to the KEY in the \bibitem in the reference list below.

\section{Introduction} \label{sec:intro}

Active regions (ARs) are areas on the Sun with strong magnetic fields where various activities take place. The ARs exhibit complex spatial and temporal behaviors from the photosphere to the corona. Studying the evolution of ARs and their relationship with solar eruptions is useful in predicting eruptive events \citep{Schrijver2007, Bobra2015, Barnes2016, Sun2022, Zhang2022} and new solar cycles \citep{Petrovay2020}, and in understanding the long-term evolution of solar activities \citep{Hathaway2010, Ravindra2022}.
Prominences are also one of the typical solar features in the solar atmosphere. They are observed as bright structures suspended over the solar limb but are seen as dark elongated structures called filaments on the solar disk \citep{Tandberg1995, Parenti2014, Vial2015, chen20}. Different from other activities in ARs which are concentrated near low latitudes, filaments also migrate toward the polar regions during each solar cycle in addition to the latitudinal migration towards the equator \citep{Li2010, Hao2015, Chatterjee2017}, which shows a close connection with the global circulation of magnetic flux that is intimately related to solar dynamo.

Prominences/filaments have been regularly observed historically on the ground by H$\alpha$ telescopes and from space in the EUV channels in recent decades (e.g., SOHO/EIT, STEREO/EUVI, and SDO/AIA). The continuous upgrade of the observational spatial and temporal resolution raises a challenge and also an opportunity to deal with and analyze huge amounts of data which are hard to check thoroughly by human inspection. A lot of automated detection methods have been developed both for solar filaments \citep[e.g.,][]{Gao2002, Shih2003, Fuller2005, Bernasconi2005, Qu2005, Yuan2011, Hao2013, Bonnin2013, Hao2015, Zhu2019} and for prominences \citep{Foullon2006, Wang2010, Labrosse2010, Loboda2015}, which can efficiently and accurately detect the morphological features and properties of the prominences/filaments. For example, \citet{Hao2015} applied their automated detection method to process the full-disk H$\alpha$  images mainly observed by Big Bear Solar Observatory during the period from 1988 to 2013, and detected 67,432 filaments in recent three solar cycles. They obtained the ``Butterfly diagram" of the filaments and conducted a quantitative analysis of the filament features such as the filament number, area, length, tilt angle, barb number, latitudinal migrating velocity, and the north-south (N-S) asymmetry of the filament features. \citet{Wang2010} developed an automated detection procedure called Solar Limb Prominence Catcher And Tracker (SLIPCAT) and applied it to the STEREO/EUVI 304  \AA \ data during 2007--2009. They obtained 9,477 prominences and got several interesting statistical results on the morphological features and vertical speeds of the prominences.

Although it is much easier to detect prominences above the solar limb than filaments on the solar disk since the background of the former is the faint corona, it is quite difficult to distinguish prominences from ARs since both are bright features \citep{Hao2018}. \citet{Foullon2006} employed multi-wavelength observations to exclude ARs when identifying prominences. \citet{Wang2010} adopted the linear discriminant analysis (LDA) based on the shape and brightness of the selected prominence candidates to exclude non-prominence features. \citet{Labrosse2010} used the method of moments combined with a Support Vector Machine (SVM)  classifier to discriminate prominences from active regions and quiet corona. \citet{Loboda2015} identified prominences based on their characteristics in He II 304 \AA. According to our experience, detecting prominences via classical image processing methods can generate a well-defined boundary of each prominence,  but the whole process relies on a series of thresholds set through a relatively large amount of trials and errors. Moreover, the robustness of these methods also suffers from being unstable due to the degradation of the detector sensitivity. The performances of graphics processing units (GPUs) and artificial neural networks (ANNs) have been advanced significantly in recent years. In particular, convolutional neural networks (CNNs) have been leading the trend of machine learning in feature segmentation for years since AlexNet \citep{krizhevsky2012} came out in 2012. Recently, CNNs have been adopted for the automated detection of filaments and are demonstrated to have relatively high performance \citep{Zhu2019, Liu2020, Liu2021, Guo2022}.

In this paper, we develop an efficient and robust automated detection method for prominences and ARs above the solar limb based on CNNs and then make a statistical analysis of their morphological features and long-term variations. Our deep learning architecture is based on U-Net \citep{Ronneberger2015}, which has an excellent performance in semantic segmentation since it can gain strong robustness even from a small data set. In Section \ref{method}, we describe our pipeline for the automated detection system and then give a general introduction to the deep learning architecture followed by a description of its performance. The statistical results of prominences and ARs detected by our method are presented in Section~\ref{results}, which are compared with the published results in Section~\ref{discussion&conclusion}, where we also draw several conclusions.

\section{Method}\label{method}

Although filaments and prominences are the same solar object, their observational characteristics are distinct due to different observational perspectives. As a result, the detection methods for them are significantly different \citep{Hao2018}. Numerous studies on the automated detection of solar filaments and prominences have been published in recent decades. Some notable works and their performance are selected and summarized in Table~\ref{tab_previous works}, where unavailable performance metrics are denoted by `N/A'. For instance, \citet{Liu2021} reported that their training required 1.2 hours for 50 epochs with 300 steps for each. However, since they applied the early-stopping method, the processing time per image is unspecified, which is also indicated as `N/A' in Table~\ref{tab_previous works}. From Table~\ref{tab_previous works}, we can see that the processing time per image of the methods based on deep learning is less than 1 second, much faster than those based on the traditional image processing methods. The accuracy of the deep learning methods is also remarkably high. Considering the fact that the relative scarcity of deep learning methods for prominence detection, we decide to adopt the U-Net model to develop an efficient method for prominence detection. As a byproduct, above-the-limb active regions can also be identified.

\floattable
\begin{longrotatetable}
\begin{deluxetable}{llllll}
\tablecaption{Collection of Methods of automated detection for filaments and prominences.\label{tab_previous works}}
\tablecolumns{6}
\tablenum{1}
\tablewidth{400pt}
\tabletypesize{\tiny}
\tablehead{
\colhead{References} & \colhead{Method} & \colhead{Target} & \colhead{Accuracy} & \colhead{Processing time} & \colhead{Hardware} \\
\colhead{} & \colhead{} & \colhead{} & \colhead{} & \colhead{(seconds per image)} & \colhead{}
}
\startdata
		\hline
		\citet{Gao2002}          & Region growing+Global thresholding                  & Filament      & N/A                              & $<$60                  & N/A                                   \\
		\citet{Shih2003}         & Region growing+Global and Local thresholding        & Filament      & Number: 1.0 (large size),        & 7.471                  & AMD Athlon@1.2GHz                     \\
		                         & +Morphological operation                            &               & Number: 0.904 (small size)       &                        &                                       \\
		\citet{Bernasconi2005}   & Region growing+Morphological operation              & Filament      & 0.72\tablenotemark{{\tiny a}}            & 120                    & N/A                                   \\
		\citet{Fuller2005}       & Region growing+Morphological operation              & Filament      & Number: 0.956\tablenotemark{{\tiny b}},  & N/A                    & N/A                                   \\
		\citet{Qu2005}           & Adaptive thresholding+Edge detection                & Filament      & Number: 0.8240\tablenotemark{{\tiny c}}, & 40.33                  & Intel Celeron@2.4GHz                  \\
		                         & +Morphological operation                            &               & Area: 0.9445                     &                        &                                       \\
		\citet{Scholl2008}       & Region growing+Thresholding+Morphological operation & Filament      & N/A                              & N/A                    & N/A                                   \\
		\citet{Joshi2010}        & Local thresholding+Size thresholding                & Filament      & Number: 0.81                     & 60                     & Unspecified CPU@3GHz                  \\
		\citet{Yuan2011}         & Adaptive thresholding+Morphological operation       & Filament      & Number: 0.96,                    & N/A                    & N/A                                   \\
		                         &                                                     &               & Area: 0.99                       &                        &                                       \\
		\citet{Hao2013, Hao2015} & Adaptive thresholding+Edge detection                & Filament      & Number: 0.85            & $<$ 10                 & Intel Core Duo@3.0GHz                 \\
		                         & +Morphological operation                            &               &                                  &                        &                                       \\
		\citet{Chatterjee2017}   & Region growing+Morphological operation              & Filament      & N/A                              & N/A                    & N/A                                   \\
		\citet{Ahmadzadeh2019}   & Mask R-CNN                                 & Filament      & IoU: 0.59                        & N/A                    & N/A                                   \\
		\citet{Zhu2019}          & U-Net                                      & Filament      & F1: 0.8944                       & 0.489\tablenotemark{{\tiny d}} & GTX 1070Ti                            \\
		\citet{Liu2021}          & U-Net                                      & Filament      & F1: 0.7710                       & N/A                    & Intel i7-7800X@3.5GHz  and RTX 2080Ti \\
		\citet{Guo2022}          & CondInst                                   & Filament      & F1: 0.8646                       & 0.458                  & RTX 2080                              \\
		\citet{Priyadarshi2023}  & Clustering                                          & Filament      & N/A                              & N/A                    & N/A                                   \\
		\citet{Foullon2006}      & Adaptive thresholding+Multi-wavelength observation  & Prominence    & N/A                              & N/A                    & N/A                                   \\
		                         & +Edge detection+Morphological operation             &               &                                  &                        &                                       \\
		\citet{Labrosse2010}     & SVM+Region growing+Morphological operation          & Prominence    & Number: 0.93\tablenotemark{{\tiny e}}    & 17                     & N/A                                   \\
		                         &                                                     & Active region &                                  &                        &                                       \\
		\citet{Wang2010}         & Region growing+Morphological operation+LDA          & Prominence    & N/A                              & 36.29                  & Intel Xeon@2.33GHz                    \\
		\citet{Loboda2015}       & Region-growing+Optimal threshold                    & Prominence    & Number: 0.913\tablenotemark{{\tiny f}}           & 2--3                   & 8-core@3.1GHz                         \\
		Ours                     & U-Net                                      & Prominence    & F1: 0.761, IoU: 0.614            & 1.277\tablenotemark{{\tiny g}} & AMD R7-5700X@3.6GHz and RTX 2070      \\
		                         &                                                     & Active region & F1: 0.845, IoU: 0.731            &                        &                                       \\
\enddata
\tablenotetext{a}{The number is not the filament detection accuracy, but the filament chirality accuracy based on the list of \citet{Pevtsov2003}. }
\tablenotetext{b}{\citet{Fuller2005} reported that their algorithm detected 1149 filaments in April 2002 (including 51 false positive ones) among 1232 manually recorded filaments.}
\tablenotetext{c}{The performance was reported by \citet{Yuan2011}. }
\tablenotetext{d}{\citet{Zhu2019} reported a training time of 49.3 minutes and the generation of 6040 samples. The processing time per-image is estimated to be 0.489 s.}
\tablenotetext{e}{\citet{Labrosse2010} provided the misclassification probability being $7\%$. So the correctly classified probability is $0.93$.}
\tablenotetext{f}{\citet{Loboda2015} reported that their algorithm tracked 46 prominences from 2009 September 10 to 20, among  which 42 were manually verified as prominences, while the remaining 4 were not.}
\tablenotetext{g}{The pre-processing required 0.603 s and the U-Net consumed 0.674 s per image.}
\end{deluxetable}
\end{longrotatetable}
\setcounter{table}{1}

\begin{figure*}[ht!]
	\centering
	\includegraphics[width=0.8\linewidth]{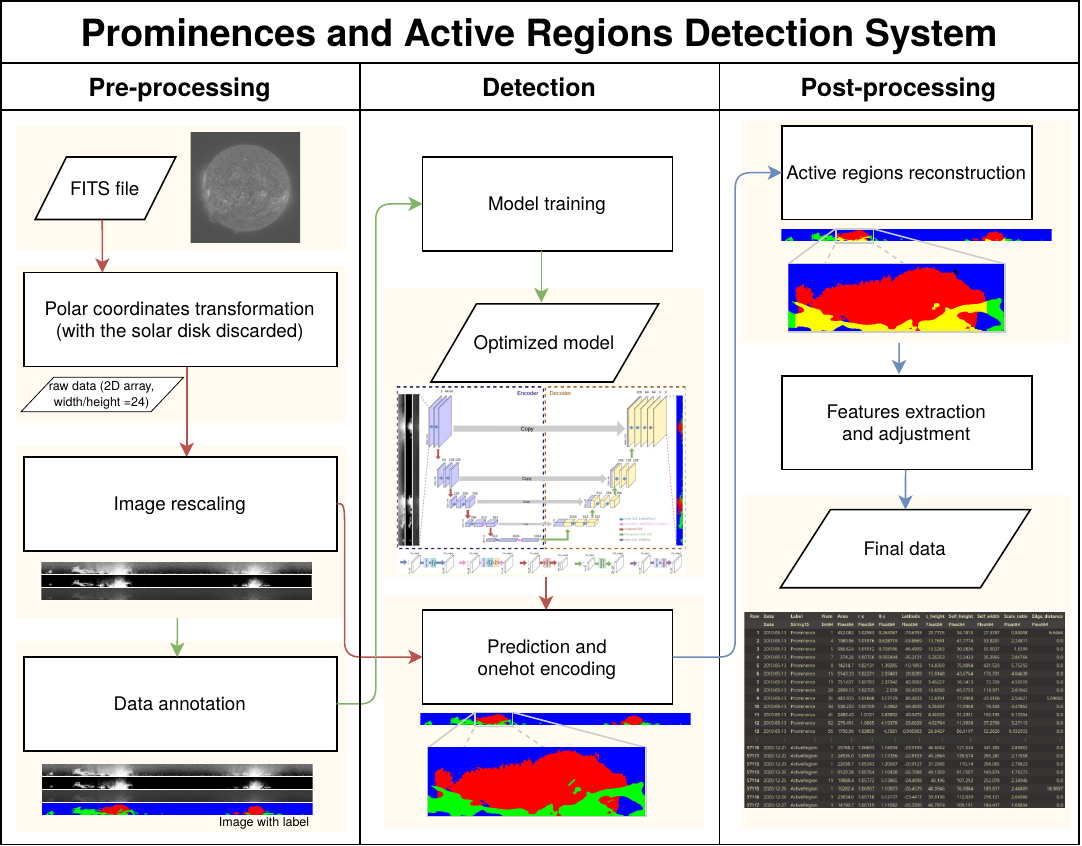}
	\caption{Flowchart showing the processes of our detection system. Objects in the same yellow box represent some functionality and its associated output or products. Red arrows link the processes of prediction. Green arrows connect the steps of the model training. Blue arrows illustrate the parts extracted from the predicted results.
	}
	\label{flowchart}
\end{figure*}

Our whole procedure is mainly divided into three modules: pre-processing, detection, and post-processing, as shown in Figure~\ref{flowchart}. In the pre-processing step, we apply several preliminary approaches to enhance and extract the off-limb images. Besides, we tag labels manually in a data subset, which is used for training, validating, and testing the deep learning module. In the detection step, we first try different parameters and train the model over and over again until we get a good enough model, and then apply it to all data. The combination of red and blue arrows shows the workflow for all data, whereas the green arrows link the workflow for model training. Since the whole pipeline does not depend on a particular order of the given images, we can parallelize the modules of our system on a computer cluster to process all the images in a shorter time. The entire automated detection system is mostly written in Julia \citep{Bezanson2017}, which is an open-source language initially designed for high performance. The deep learning module in our pipeline is constructed with PyTorch \citep{NEURIPS2019_9015}. The following subsections explain the details of each module.

\subsection{Data Acquisition}\label{data}

Prominences are routinely observed from space in He II 304 \AA\ in the past decades \citep{Labrosse2010, Wang2010}. For this study, we select the full-disk  304 \AA\ images observed by the Atmospheric Imaging Assembly (AIA) onboard Solar Dynamics Observatory  (SDO; \citealt{Lemen2012}). The SDO/AIA has been acquiring 304 \AA\ images in a 12 s cadence with a pixel size of $0\farcs6$ (each image has $4096 \times 4096$ pixels) since its launch on 2010 February 11. The data cover almost the whole solar cycle 24, which allows us to analyze the long-term variation. For every single day from 2010 May 13 to 2020 December 31, we select one image per day for detecting prominences and ARs.

\subsection{Pre-processing}\label{pre-process}

\begin{figure*}[ht!]
	\centering
	\includegraphics[width=\linewidth]{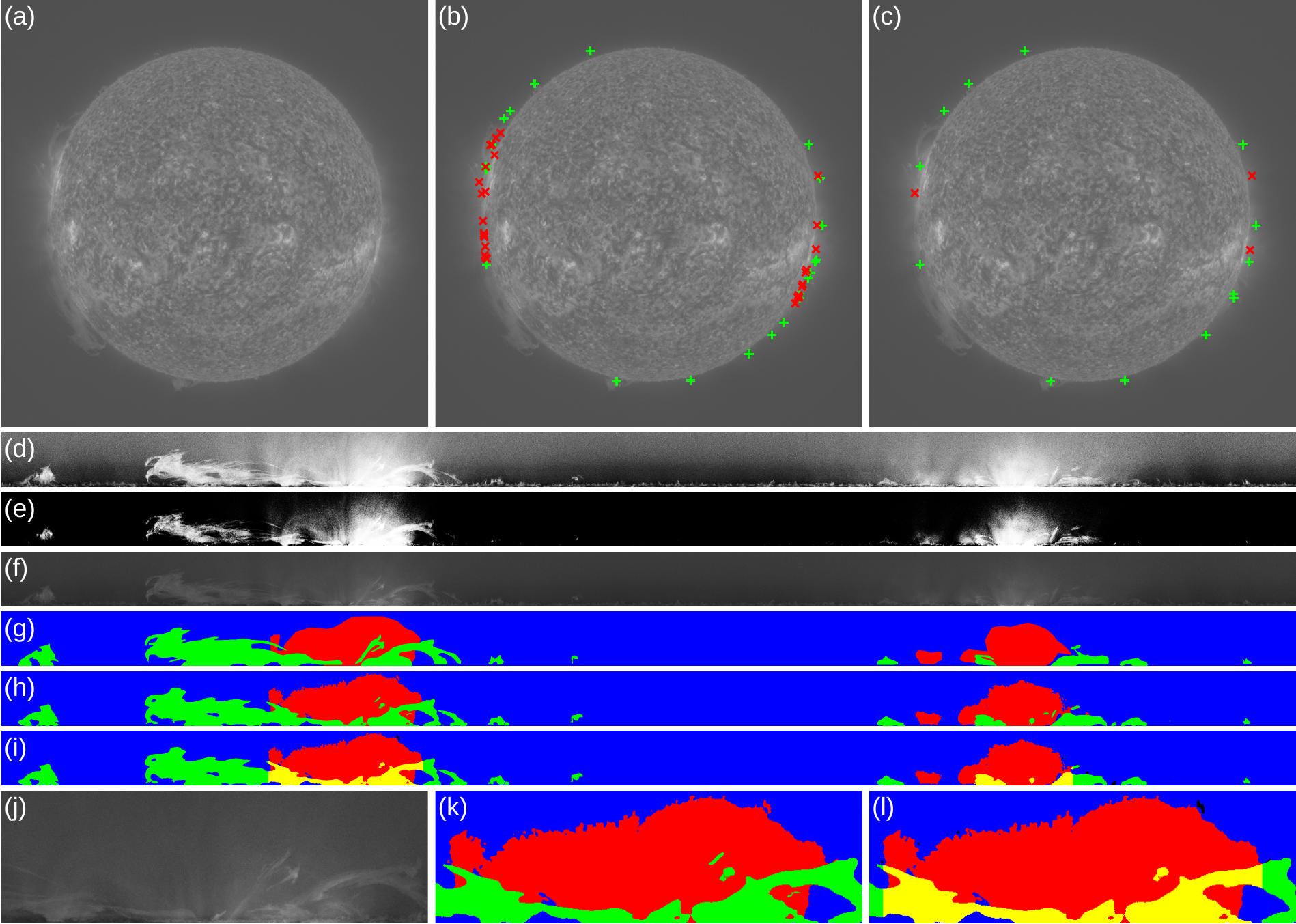}
	\caption{An observed 304 \AA \ image obtained on 2014 June 1 as an example to show the detection modules. (a) The raw data after a logarithmic rescaling. (b) The predicted prominence and AR candidates directly from U-Net are marked as green ``$+$"   and red ``$\times$", respectively. (c) The final prominences and ARs after the post-processing. The symbols have the same mean as (b).  (d--f) The ring-shaped off-limb regions are transformed into rectangles in polar coordinates with different rescaling methods. (g) The manually labeled prominences (green), ARs (red), and background (blue). (h) The predicted prominence and AR candidates by our U-Net model.  (i) The final detected results after the post-processing. The yellow blocks stand for the intersection between prominences and ARs. Panels (j--l) are the enlarged parts of (f), (h), and (i), giving the zoomed details. }
	\label{flowchart_example}
\end{figure*}

As illustrated in Figure~\ref{flowchart}, we first conduct pre-processing before using the detection module. In order to save computation resources and detect prominences and ARs more efficiently, the original image is transformed to the polar coordinates with the solar disk discarded. The transformed image becomes rectangular with the horizontal coordinate representing the position angle measured clockwise from the south pole in degrees and the vertical coordinate representing the heliocentric distance in pixels. Then we perform image rescaling to make the prominences and ARs more distinguishable from the background in the transformed image as done by \citet{Wang2010}, where each pixel is divided by the average intensity of all the pixels at the same height. The deep learning model we adopt is a supervised machine learning method, which means we first have to provide the data sets with prominences, ARs, and backgrounds labeled. However, the current approach can not yet generate specific boundaries of these features manually. It is hard to label the boundaries by manual inspection such as the boundaries of ARs, and the overlap parts between prominences and ARs. In the next step, we rescale the image in three different ways to solve this problem. For the first one, we adopt a technique called tone-mapping to enhance the contrast. A simple useful tone-mapping is the Reinhard mapping \citep{Reinhard2002}. We adopt a similar mapping:
\begin{equation}
	f(x) = \frac{x^5}{1+x^5}\,\,,
\end{equation}
where $x$ is the intensity of each pixel and $f(x)$ is the rescaled intensity of each pixel. After that, we are able to maintain most information of the image and achieve high contrast between prominences, ARs, and the background.
For the second one, we first adopt another mapping:
\begin{equation}
	f(x) = \frac{x^{10}}{1+x^{10}}.
\end{equation}
With this mapping, despite weak solar activity during the declining phase of the solar cycle, the processed images maintain a high contrast. Then we use the other tone mapping as done by \citet{durand2002fast} to achieve the higher contrasts between ARs boundaries and the background. However, for active region prominences, they are rooted in ARs and are not bright enough to be easily distinguished from background AR, i.e., there are overlapping parts between prominences and ARs. Thus we provide the third way using the following mapping:
\begin{equation}
	 f(x) = \log(1+x)\,\,.
\end{equation}
Besides, we adopt these three kinds of tone-mapped images to generate a tensor that has three channels as our model input since more constraints can make the model training faster and more accurate. For all of the three channels, we linearly map the tone-mapped image to $[0, 1]$ respectively. As an example, Figure~\ref{flowchart_example}(a) shows the 304 \AA \ raw image with a logarithmic rescaling, and panels (d--f) show the annular off-limb corona in the Cartesian coordinates from the polar coordinates with the three rescaling methods mentioned above. The long side of the Cartesian map is 24 times the short side in pixels.  Considering hardware requirements and time cost, we resize the image to 4608 $\times$ 192 pixels, about a quarter of their original size.

Here we manually label 128 monthly images from 2010 June 1 to 2021 January 1, for convenience and in order to get a uniform distribution, all on the first day of each month. We divide the 128 pairs into three groups for training, validating, and testing the model. In more detail, 21 pairs observed on March 1 and September 1 each year are selected as the validation data, and 22 pairs obtained on June 1 and December 1 each year are chosen to be used for testing. The remaining 85 pairs are used for training our model. There is a simple way called data augmentation to enlarge the existing data set by performing a specific transform to the images \citep{Miko2018}. Generally speaking, we can rotate an image by small angles, flip it horizontally or vertically, zoom in and out at small scales, or combine these operations to obtain different training sets. Although the images obtained by these transformations are not very different from the original ones, it can reduce the risk of overfitting since the convolution kernel can be invariant to translation, rotation, or a combination of these. We use data augmentation in our case to increase the amount of data.

\subsection{Deep Learning Architecture for Detection}\label{dl architecture}

\begin{figure*}[ht!]
	\centering
	\includegraphics[width=\linewidth]{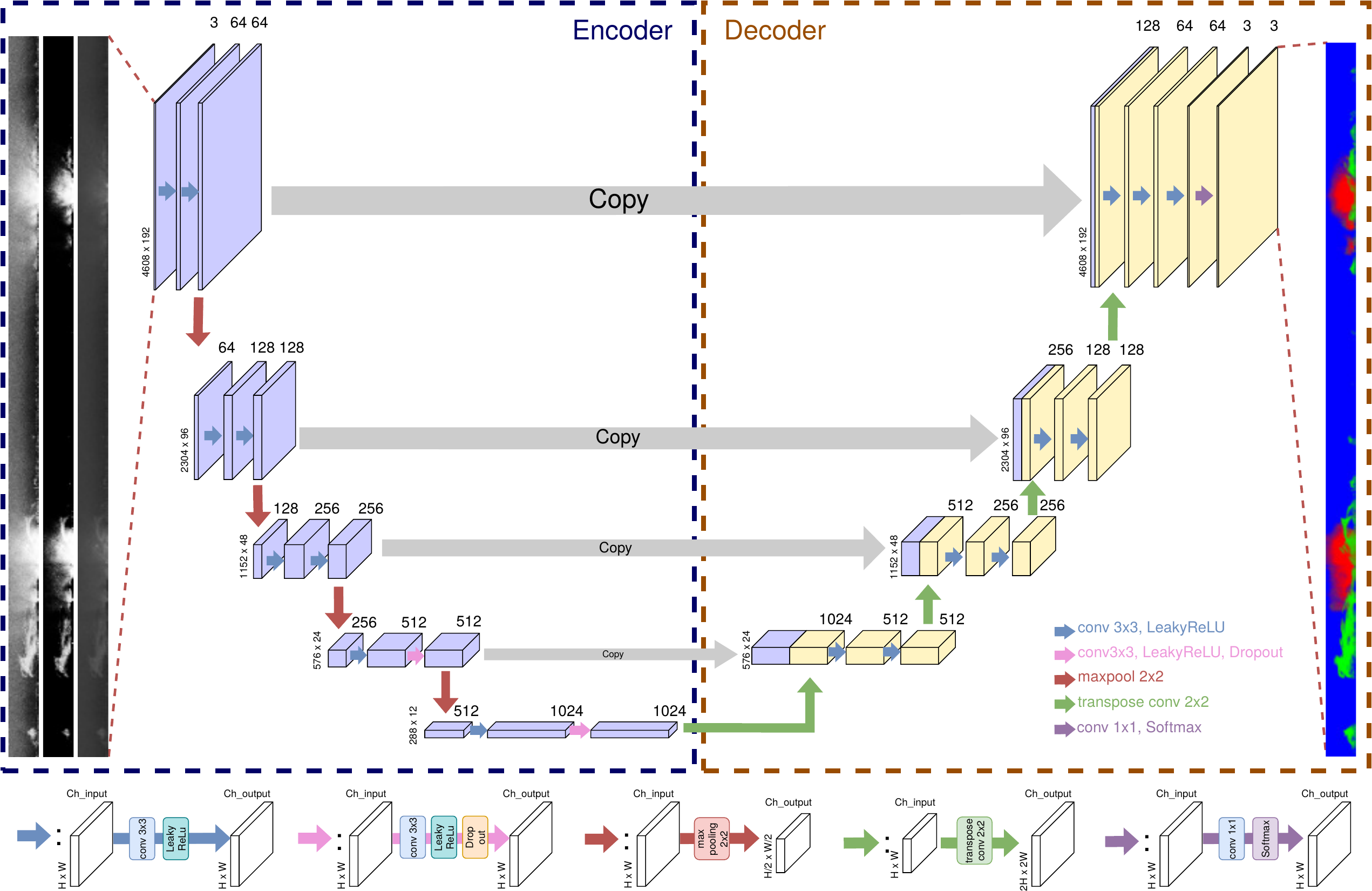}
	\caption{Schematic representation of our U-Net architecture. It consists of an Encoder (left blue box) and a Decoder (right red box). The leftmost input is a 3D array combing three tone-mapped data, and the rightmost output is the prediction 3D array: prominences (green), ARs (red) and background (blue). The cubes represent feature maps, where the dimension of each map is indicated on its left, and the number of channels is indicated above it. Operations for each channel are represented by arrows: the blue arrows indicate a convolution operation with a $3 \times 3$ kernel followed by a LeakReLu activation function;  the pink arrows indicate a DropOut operation in addition to the operation indicated by the blue arrow; the red arrows represent a down-sampling operation using max pooling with $2 \times 2$ pooling size; the green arrows indicate an up-sampling operation by transpose convolution operation with a $3 \times 3$ kernel; the gray arrows indicate the array in Encoder splice to the array in Decoder. H and W refer to the height and width of each feature map, respectively.
	}
	\label{unet}
\end{figure*}

Machine learning is the study of algorithms that can let the machine itself ``learn" from experience. Among various kinds of machine learning methods, deep learning or neural network is a powerful set of techniques that drives innovations in different areas such as computer vision, natural language processing, biomedicine, and also in astronomy in the recent decade \citep{Smith2023, Asensio2023}.
Our method is based on the U-Net, a class of convolutional neural networks (CNNs) as shown in Figure~\ref{unet}, which can use a very limited quantity of training images to obtain more accurate recognition results \citep{Ronneberger2015}.  CNN is a powerful family of neural networks that are designed for processing image data. They tend to be computationally efficient since they require fewer parameters and can be easily parallelized on GPUs \citep{Zhang2023}. A typical CNN contains a bunch of convolutional layers, pooling layers, and dropout layers \citep{Hinton2012}, and these layers are connected by activation functions.
A convolutional layer, also known as a convolution kernel or filter, is similar to a filter in traditional image processing. It has the same purpose that works as a feature detector to get the feature maps. However, its parameters are learned rather than being set manually based on experience. In addition, the input and output in each layer have multiple corresponding channels. Intuitively, we can think of each channel as a response to a different feature. As the layers of the neural network go deeper, the dimensionality of the output channels gets larger. For example, the first input layer has 3 channels and the second layer has 64 channels as shown in Figure~\ref{unet}. Pooling layers play a critical role in compressing the size of feature maps and reducing the complexity of deep learning models while preserving important features and relationships in the input images. We adopt max pooling in our model. The output would be a feature map containing the most prominent features of the previous feature map. LeakyReLu and Softmax are two commonly used activation functions, which are defined as follows:
\begin{equation}
	\mathrm{LeakyReLu}(x; \alpha) = \max\{0,x\} + \alpha\min\{0, x\},
\end{equation}
\begin{equation}
	\mathrm{Softmax}(\mathbf{x})_i = \frac{e^{\mathbf{x}_i}}{\sum_{j=1}^{n}e^{\mathbf{x}_j}},
\end{equation}
where $n$ means the number of classes and $\mathrm{Softmax}(\mathbf{x})_i$ represents the predicted probability of the output being class $i$. We use LeakyReLu as the activation function after each convolutional layer except the last one where we use Softmax for classifying three classes, i.e., prominences, ARs, and the background.

A U-Net consists of two parts: the encoder (left side) and decoder (right side) as shown in Figure~\ref{unet}, which are also known as the contracting path and expansive path. Its architecture gets an overall U-shape. The encoder gradually compresses the input information into a lower-dimensional representation, i.e., reduces the spatial information while increasing the feature information. Then the decoder combines the feature and spatial information through a series of transpose-convolutions to return to the original image dimension.  The encoder and decoder in a U-Net are divided into several blocks. The inputs are three grayscale maps combined together as a three-channel 3D array. The output is also a 3D array that contains the predicted prominences (green), ARs (red), and background (blue). The cubes are the intermediate hidden layers representing feature maps, where the dimension of each map is indicated on its left side, and the number of channels is indicated above it. The colored arrows represent different operations. The blue arrows in Figure~\ref{unet} indicate a convolution operation with a $3 \times 3$ kernel followed by a LeakyReLU activation function. The pink arrows indicate the operations by the blue arrows in addition to a dropout operation to randomly deactivate some nodes, which are used in the deeper layers of the network to avoid overfitting. The red arrow indicates the max pooling operation with a $2 \times 2$ pooling size, which is conducted to reduce each feature map to its quarter size. Feature channels were doubled to retain pertinent information from the previous feature map. The green arrow indicates a transpose convolution with a $2 \times 2$ kernel, which applies the inverse-convolution instead of the max-pooling operation to reconstruct a new feature map with half-channels and 4 times size in the decoding part. One of the salient features of the U-Net architecture is the skip connections presented by the gray arrows as shown in Figure~\ref{unet}, which enable the flow of information from the encoder side to the decoder side, facilitating the model to make better segmentations. The final output feature map is obtained by a convolution operation with $1 \times 1$ kernel followed by the sigmoid activation function, and the combined operation is indicated by the purple arrow.

	Typically, with neural networks, we seek to minimize the error between the predicted results and the labeled ground truth. The objective function is often referred to as a loss function. It defines an objective by which the performance of the model is evaluated. The parameters learned by the model are determined by minimizing a chosen loss function. We choose the focal loss \citep{Lin2017} as our loss function, which is defined as
	\begin{equation}
		\mathrm{FL}(\hat{\mathbf{y}}, \mathbf{y}; \mathbf{w}, \gamma) = -\sum_{i=1}^{n}\mathbf{w}_i \mathbf{y}_i(1-\hat{\mathbf{y}}_i)^\gamma \log \hat{\mathbf{y}}_i,
	\end{equation}
	where $n$ represents the number of classes, $\hat{\mathbf{y}}_i$ stands for the predicted probability of class $i$, and $\mathbf{y}_i$ represents the ground truth. $\gamma$ is usually chosen to be $2$ and $\mathbf{w}_i$ stands for the weight of the $i$th class. In our model, there are three classes, i.e., prominence, AR, and background. Focal loss benefits when the samples of different classes are imbalanced. It is very important to tweak the weights of the model during the training process, to make our predictions as correct and optimized as possible. The optimizer ties together the loss function and model parameters by updating the model in response to the output of the loss function. We choose ADAM \citep{Kingma2014} as the optimizer in our model.  A dropout operation is often used to avoid overfitting during the training. It either sets an element to $0$ with a probability $0<p<1$, or multiplies $\frac{1}{1-p}$ to the element with a probability $1-p$. The parameters in our model are listed in Table~\ref{tab:paras}.

\begin{table}[ht!]
	\centering
	\caption{Parameters used in our U-Net model. }
	\begin{tabular}{ll}
		\hline
		Convolution Kernels              & See Figure~\ref{unet} \\
		Active functions                 & See Figure~\ref{unet} \\
		Optimizer                        & ADAM                  \\
		Epochs                           & 100                   \\
		Batchsize                        & 1                     \\
		Learning rate                    & $10^{-4}$             \\
		L2 regularization                & $10^{-6}$             \\
		Dropout rate                     & $20\%$                \\
		$\alpha$ in LeakyReLu            & $0.1$                 \\
		$\mathbf{w}_{\text{AR}}$         & $2.5$                 \\
		$\mathbf{w}_{\text{prominence}}$ & $4.0$                 \\
		$\mathbf{w}_{\text{background}}$ & $1.0$                 \\
		$\gamma$ in Focal Loss           & $3.0$                 \\
		\hline
	\end{tabular}
	\label{tab:paras}
\end{table}

\subsection{Post-processing}
By applying the trained U-Net to the AIA 304 \AA\ data, we get the prominence and AR candidates. An example is shown in panels (h) and (k) of Figure~\ref{flowchart_example}. Two cases are considered in the post-processing step before the final feature identification. In the first case, one type of prominences, i.e., active-region prominences, are usually rooted in ARs, so that these prominences and active regions may overlap with each other, as shown in Figure~\ref{flowchart_example}(k). The overlapping part not only belongs to the prominence but also to the AR at the same time. However, these structures are usually detected as prominences. Thus we take the following steps to add a label to the AR. First, we identify the minimum bounding rectangle of an AR. Then, we check if there is any prominence overlapping with this AR. If so, the intersection of the prominence and the rectangle is also recognized as a part of the AR. In the second case, some fluctuations in the raw data may yield small regions which are recognized as prominences or ARs since our model is very sensitive, even though it is hard to distinguish whether these regions are prominences or ARs manually. Therefore, we perform a filter on the area to avoid this situation: any identified prominences smaller than  $250~\text{Mm}^2$ and ARs smaller than $500~\text{Mm}^2$ are discarded in our results. In addition, we consider the effect of the error due to the transformation from the annular shape to a rectangle on the extracted features, which makes the results more precise and reliable. An example of the post-processing is shown in Figure~\ref{flowchart_example}. Figures~\ref{flowchart_example}(g--i) show the manually labeled image, the predicted prominence and AR candidates by our U-Net model, and the final identification results after the post-processing, respectively. The green, red, and yellow-shaded regions in panels (g--i) represent the predicted prominences, ARs, and the intersection between prominences and ARs, respectively. In order to see them more clearly, we enlarge the prominence, ARs, and the intersection parts of panels (f--h), and (i) in panels (j--l), respectively. Figures~\ref{flowchart_example}(b) and (c) show the detected prominence and AR candidates by our U-Net model and the final prominences and ARs after post-processing, respectively.

\subsection{Performance}\label{method - performance}
\begin{figure}[ht!]
	\centering
	\includegraphics[width=\linewidth]{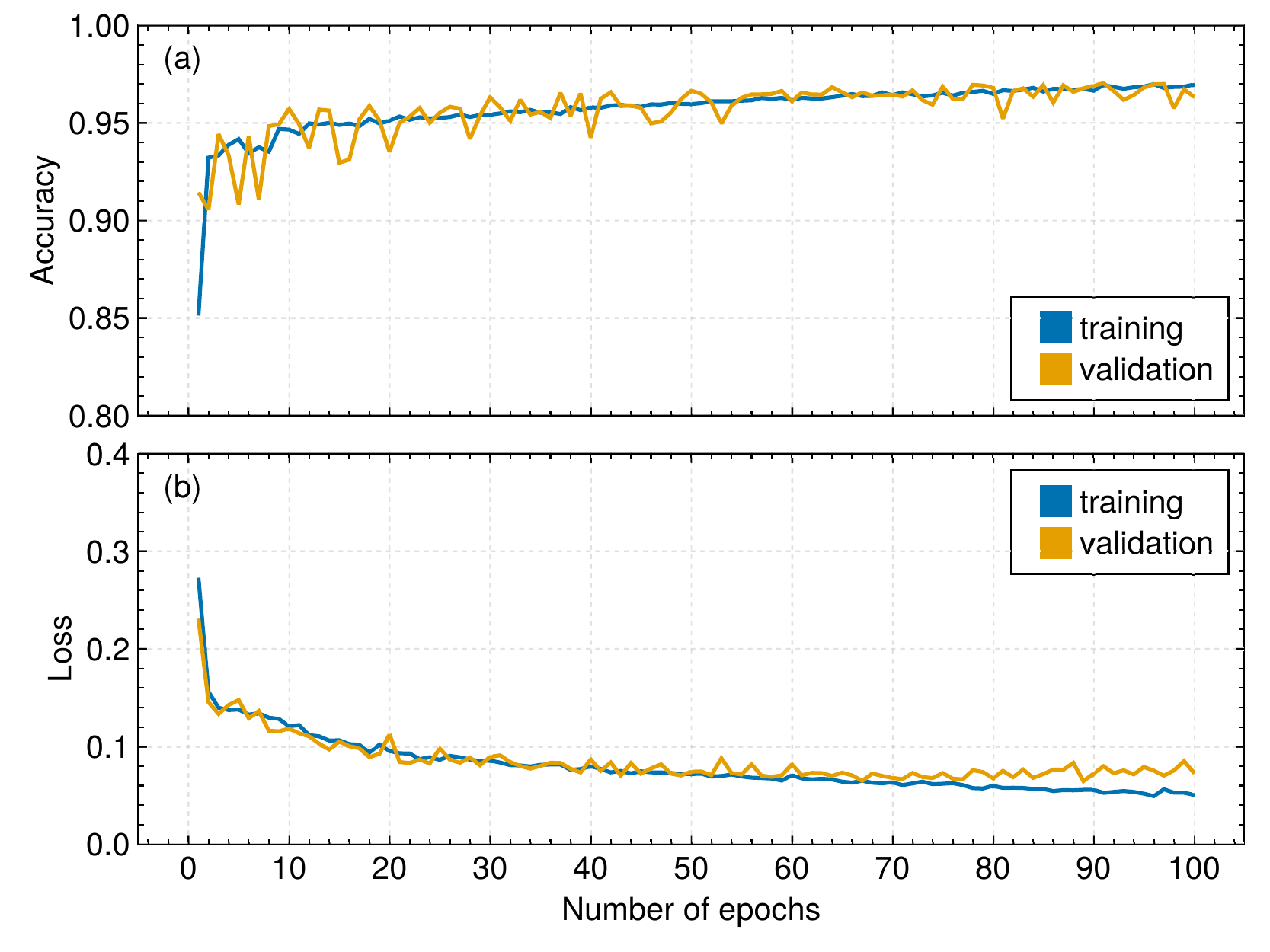}
	\caption{Accuracy and loss during training. }
	\label{acc_loss}
\end{figure}

We record the accuracy and loss during the process of training, as shown in Figure~\ref{acc_loss}. The blue and orange lines stand for the training and validating processes respectively. As seen from  Figure~\ref{acc_loss}(b), the loss on the training data becomes lower than that of the validation data since epoch 60, which means the model is probably overfitted. As a result, we choose the model that has the minimal loss of validation data after an epoch of training, which is the 89th model during training.

There are three more metrics we are interested in: the precision $P$,  recall ratio $R$, and $F1$ score, which are defined as
\begin{eqnarray}
	P = \frac{TP}{TP + FP}\,,
\end{eqnarray}
\begin{eqnarray}
	R = \frac{TP}{TP + FN}\,,
\end{eqnarray}
\begin{eqnarray}
	F1 = \frac{2TP}{2TP + FP + FN} = \frac{2 R P}{R + P}\,,
\end{eqnarray}
where $TP$, $FP$, and $FN$ denote the numbers of true positive, false positive, and false negative cases, respectively.
The precision, recall ratios, and F1 score of our new U-Net method with the ground truth on 22 testing examples are listed in Table~\ref{tab:perform}. Basically, it shows that the performance of our U-Net method has a good detection performance for both prominences and ARs. For ARs,  the precision and recall ratios of our model are substantially higher. It probably benefits from the second channel of our input, and one example is shown in Figure~\ref{flowchart_example}(e). Additionally, it indicates that improving the pre-processing method could advance the performance of the deep learning model.

\begin{table}[ht!]
	\centering
		\caption{Performance metrics of our U-Net method.\label{tab:perform}}
		\begin{tabular}{lllllll}
			\hline
			      \multicolumn{3}{c}{Active Regions}	& \multicolumn{3}{c}{Prominences}                                 \\
			      P		& R			& F1			& P			& R 			& F1      	\\ \hline
			 $81.2\%$	& $88.0\%$	& $0.845$	& $66.8\%$	& $88.4\%$	& $0.761$ \\ \hline
		\end{tabular}
\end{table}

\begin{figure}[ht!]
	\centering
	\includegraphics[width=\linewidth]{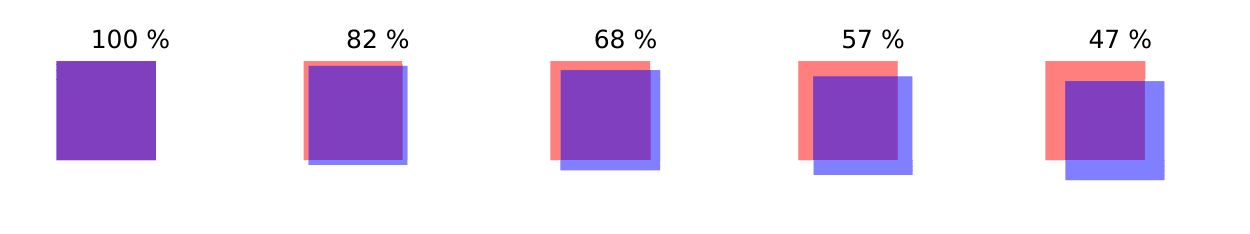}
	\caption{Examples of the IoU. The blue block and red block represent two sets and the purple region represents where they intersect. }
	\label{IoU_example}
\end{figure}

Besides, in semantic segmentation tasks, the intersection over union (IoU), also known as the Jaccard index, is often used to evaluate the performance of a model \citep{Rezatofighi2019}. For two sets $A$ and $B$, the IoU of them is defined as
\begin{equation}
	\text{IoU}(A, B)=\frac{|A \bigcap B|}{|A \bigcup B|}\,\,.
\end{equation}
It is easy to prove that the IoU between the precision $P$ and the recall ratio $R$ is
\begin{equation}
	\text{IoU} = \frac{RP}{R+P-RP}\,\,.
\end{equation}
Figure~\ref{IoU_example} gives examples of the IoU, and in most cases, over $60\%$ is a good enough score.

\begin{figure}[ht!]
	\centering
	\includegraphics[width=\linewidth]{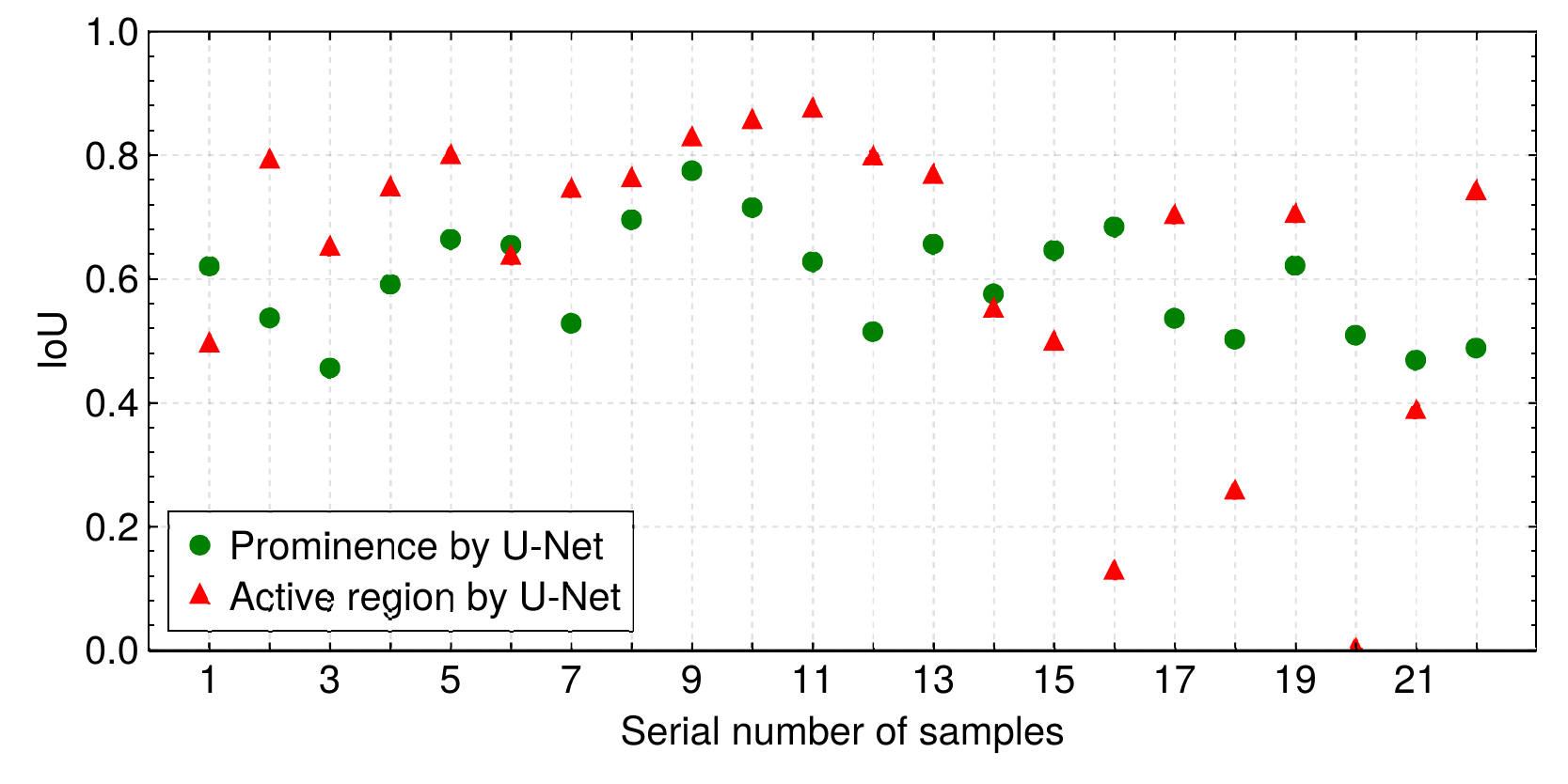}
	\caption{IoU metrics of prominences and ARs by U-Net.
	}
	\label{iou}
\end{figure}

The IoU metrics of our 22 testing examples are displayed in Figure~\ref{iou}, which indicate our U-Net method provides efficient segmentation results. We also find that ARs are more detectable than prominences, which may be due to the ground truth data set used for the analysis.

In terms of processing time for each image, the pre-processing required 0.603 seconds and the U-Net consumed 0.674 seconds, respectively. Note that our code was tested on an ordinary desktop computer (CPU: AMD R7--5700X 3.6GHz, GPU: NVIDIA GeForce RTX 2070). The processing speed will be much faster if a more capable running environment is used since the code is parallelized on GPUs. Our U-net method and the testing examples are available on GitHub \footnote{\url{https://github.com/tianmaitianmai/SPARDA}}.

\section{Result}\label{results}

As mentioned in Section~\ref{data}, we select one image per day for detecting the prominences and ARs from 2010 May 13 to 2020 December 31. In total $3,865$ images were processed. The butterfly diagrams of the prominences and ARs are obtained. We present the statistical results of the prominence and AR features in the following subsections, respectively.

\subsection{Prominences}

\subsubsection{Distributions of the Prominence Number}

\begin{figure}[ht!]
	\centering
	\includegraphics[width=\linewidth]{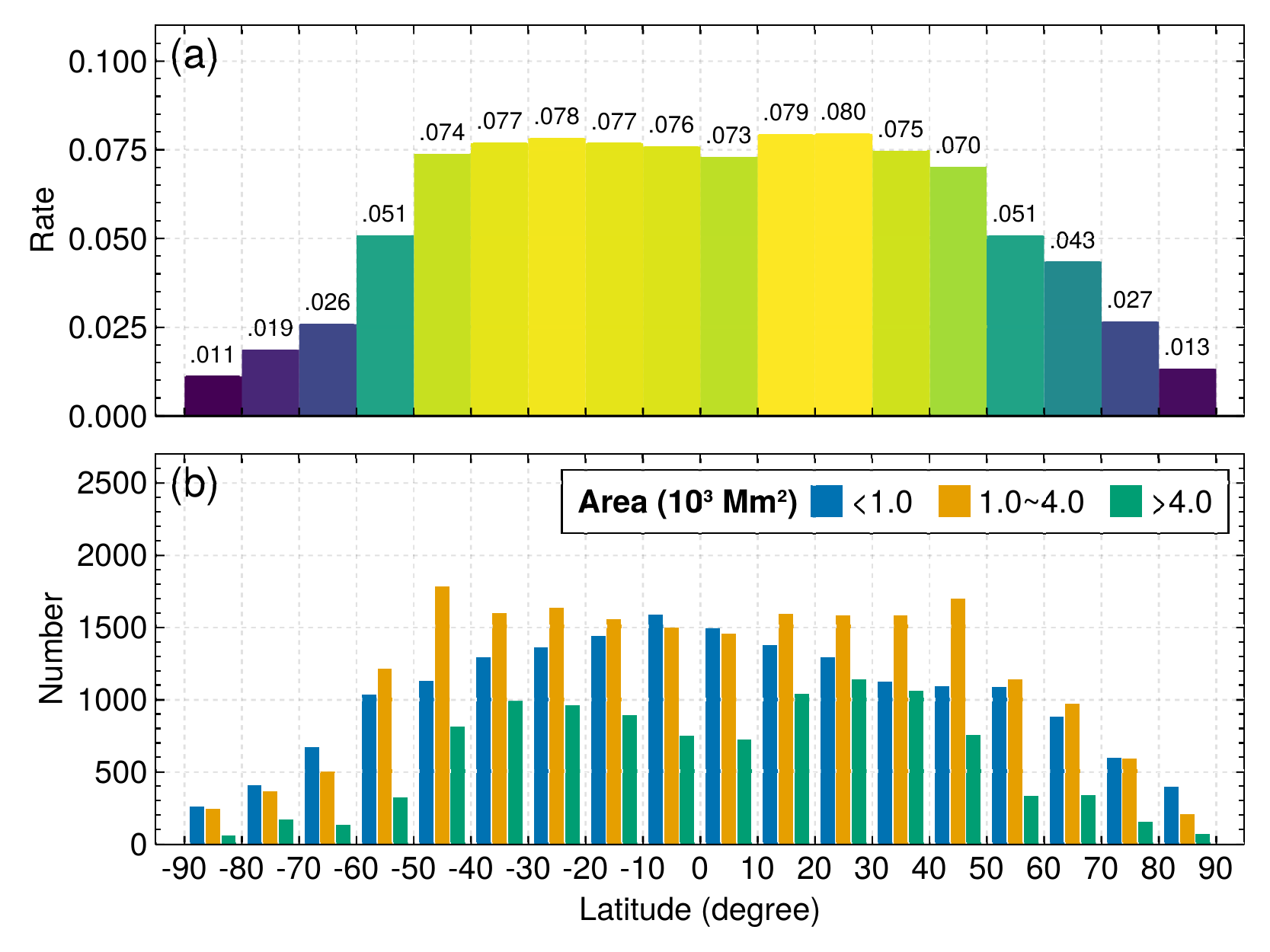}
	\caption{Distribution of the prominence number with respect to latitude. (a) The normalized prominence number in different latitude bands. (b) Distribution of the prominence number within three area groups in different latitude bands.}
	\label{p_number_latitude}
\end{figure}

\begin{figure}[ht!]
	\centering
	\includegraphics[width=\linewidth]{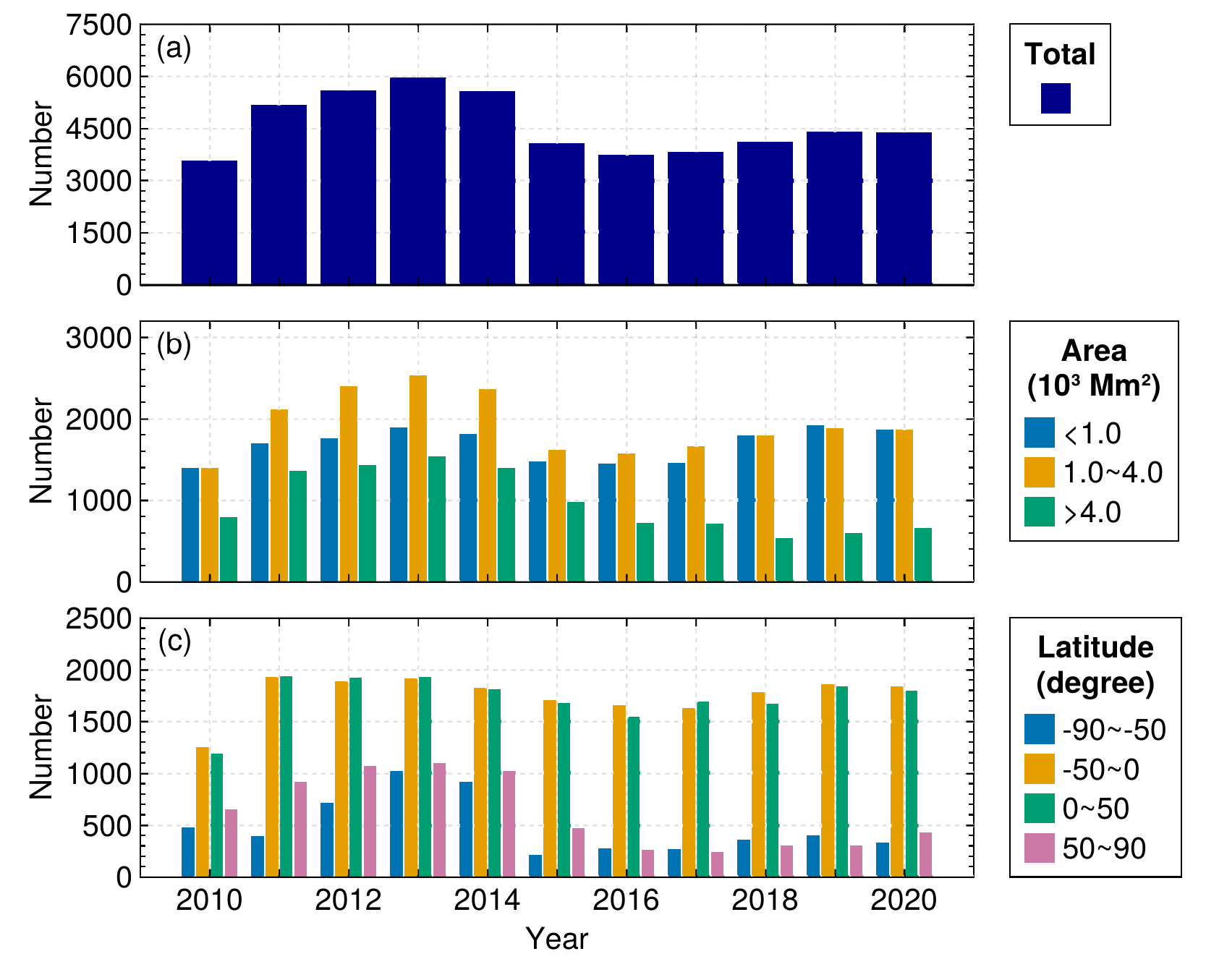}
	\caption{Distribution of the prominence number from 2010 to 2020. (a) The total prominence number in each year. Panels (b) and (c) are similar to pane (a) but for the number of prominences that are grouped by areas and latitude bands, respectively.}
	\label{p_number_year}
\end{figure}

We detected $50,456$ prominences from 2010 May 13 to 2020 December 31. Note that the number corresponds to the sum of all detected prominences, i.e., a prominence might be counted several times when it remains visible for several days above the solar limb. The latitudinal distribution of the prominences is plotted in Figure~\ref{p_number_latitude}, which shows a bimodal distribution. As indicated by Figure~\ref{p_number_latitude}(a), the number of prominences has peaks in the latitude band $20^{\circ}$--$30^{\circ}$ in both hemispheres, though there is no significant difference among the bands within the latitude range of $0^{\circ}$--$50^{\circ}$. In order to analyze the characteristics of the prominence size, the detected prominences are divided into three groups based on their areas, i.e.,  $<$$1000\ \text{Mm}^2$, 1000--4000 $\text{Mm}^2$, and $>$$4000\ \text{Mm}^2$, with the ratio being about $5:6:3$. Figure~\ref{p_number_latitude}(b) shows the latitudinal distribution of the prominence number in the three area ranges.  The distribution of the prominences with areas over $4000\ \text{Mm}^2$ is similar to the total number distribution in Figure~\ref{p_number_latitude}(a), i.e., both are symmetrically bimodal. The distribution of the prominences with areas in the range 1000--4000 $\text{Mm}^2$ is also bimodal but with peaks at higher latitude band, $40^{\circ}$--$50^{\circ}$. In contrast, the distribution of the number of the prominences with areas smaller than $1000\ \text{Mm}^2$ is not bimodal. It has a peak around the equator. This might imply that the formation of the smallest prominences is slightly different from larger prominences.

Figure~\ref{p_number_year} plots the yearly distributions of the prominence numbers within solar cycle 24. The total number of prominences rises from the year 2010 and reaches a peak around the year 2013, then decreases in the following years and reaches the minimum in the year 2016. After that, it rises again slightly in the rest of the declining phase, as shown in Figure~\ref{p_number_year}(a). From Figure~\ref{p_number_year}(b),  we can see that the numbers of the three groups of prominences with different areas are not much different. However, in the declining phase from 2016 to 2020, the number of the largest prominences with areas over $4000\ \text{Mm}^2$ gradually decreases to nearly half that in the solar maximum. In contrast, the numbers of the prominences with areas smaller than $4000\ \text{Mm}^2$ decrease first and increase again, becoming slightly less than in the solar maximum. These results indicate that only the largest prominences follow the sunspot cycle and there are still many smaller prominences even when the sunspot number approaches the minimum in solar cycle 24.

With the latitude $50^\circ$ being the boundary, we divide the prominences into high/low latitude prominences in both hemispheres in the following analysis since the normal solar activity is usually defined as events at latitudes below $50^{\circ}$ \citep{Sakurai1998, Li2008, Hao2015, Diercke2019}. It is found that about $76\%$ of the prominences are low-latitude ones. The yearly distributions of the prominence numbers in high and low latitude bands are displayed in Figure~\ref{p_number_year}(c). The low-latitude prominences present another peak in 2019 near the solar minimum, in addition to the main peak in the solar maximum, indicating that they do not follow the solar cycle as well. In contrast, the number of high-latitude prominences peaks around the solar maximum and then decreases significantly in the declining phase of the solar cycle. It is also noted that the number of high-latitude prominences remains roughly unchanged in the whole declining phase.

As for the hemispheric asymmetry, it is seen that the low-latitude prominences are comparable in number between the two hemispheres during the entire solar cycle. On the other hand, the northern hemisphere is slightly dominant in high-latitude prominences from 2010 to 2015. Thereafter, the two hemispheres are even.

\subsubsection{Butterfly Diagram of Prominences}

	\begin{figure}[ht!]
		\centering
		\includegraphics[width=0.99\linewidth]{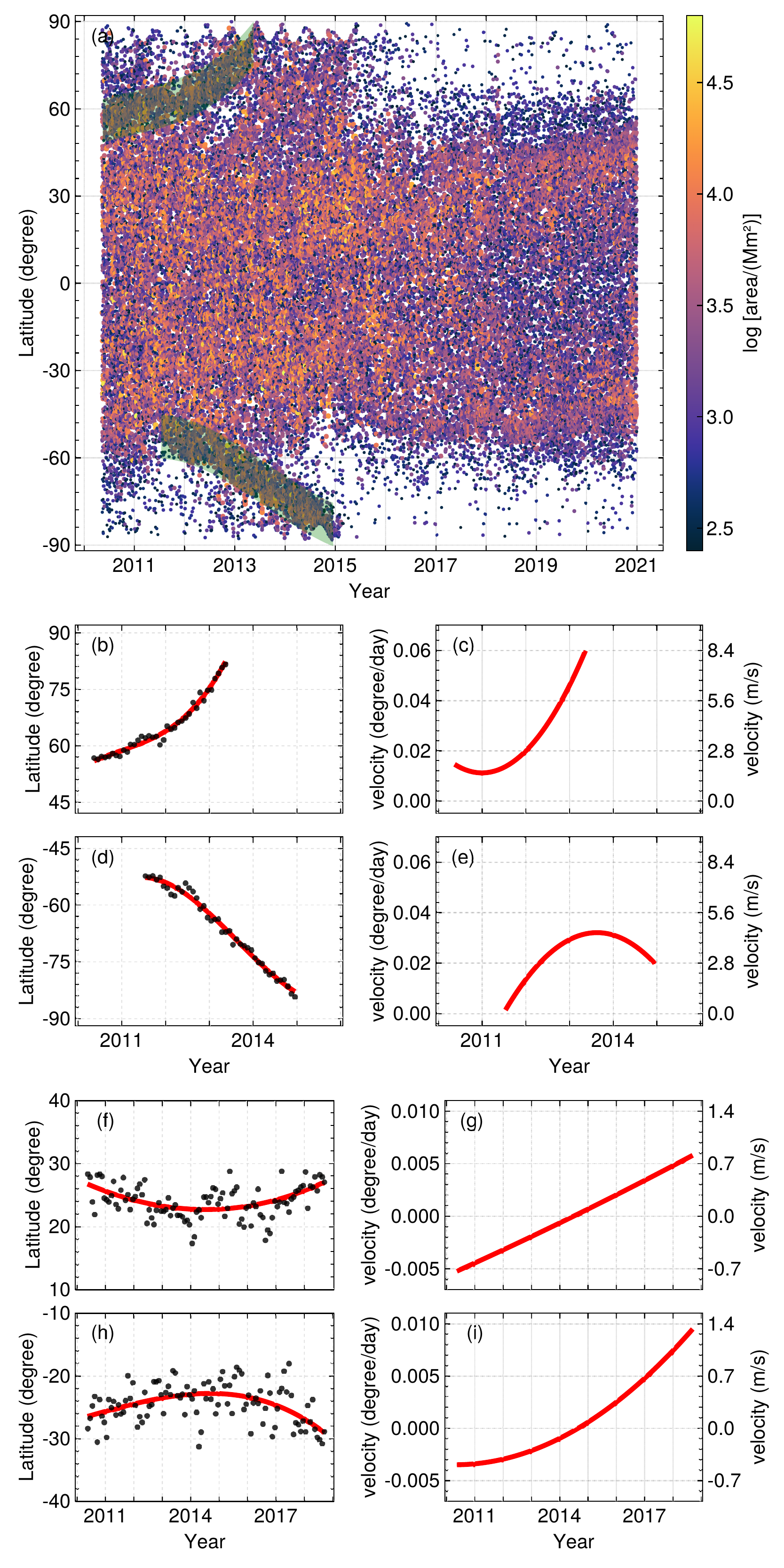}
		\caption{Butterfly diagram and latitudinal migration of prominences. (a) Butterfly diagram of the prominences from 2010 to 2020. Each dot represents a single prominence. The colorbar shows the logarithm of the prominence areas. The green bands show the data used to calculate the drift velocities. (b) Temporal evolution of the monthly average latitude of the high-latitude prominences in the northern hemisphere, which is fitted with a cubic polynomial function indicated by the red line. (c) Fitted drift velocity variations of the high-latitude prominences in the northern hemisphere. Panels (d--e) are similar to panels (b) and (c) but for the high-latitude prominences in the southern hemisphere. Panels (f--i) are similar to panels (b--e) but for the low-latitude prominences.}
		\label{butterfly_and_driftV_total}
	\end{figure}

The equatorward latitudinal migration of sunspots, which forms the butterfly diagram, reflects the meridional flow related to the solar dynamo  \citep{Hathaway2010, chou21}, and so do the low-latitude filaments \citep{Hao2015}. The time--latitude diagram of the prominences is plotted in Figure~\ref{butterfly_and_driftV_total}(a). Each dot in the figure represents a single prominence, where the latitude corresponds to the geometric center of the prominence. It is seen that the latitude distribution of the prominences is different from sunspots in several aspects. First, prominences are more dispersive than sunspots. Second, the low-latitude prominences migrate toward the equator from 2010 to 2014, similar to sunspots, but then turn around and migrate toward higher latitudes after 2014. Third, in addition to the low-latitude prominences, there are many high-latitude prominences, which migrate toward the pole in each hemisphere. Such migration of high-latitude prominences toward the polar region is the well-known ``rush to the poles" \citep{Hyder1965, Shimojo2006, Li2010, Gopalswamy2016, Xu2018, Tlatova2020}.

Since the latitude distribution is dispersed, we take the monthly average latitudes of the prominences in both hemispheres to analyze the latitudinal migration and drift velocities quantitatively. The least squares fitting with a cubic polynomial function is adopted to fit the data. Here, we first set a preliminary latitude band and fit the monthly average latitudes of the prominences to get $c^0(t)$. Then we reset the latitude bands to $[c^0(t)-\delta, c^0(t)+\delta]$ and fit the data again to get $c^1(t)$. The value of $\delta$ is set to $7.5^{\circ}$ for prominences. These iterations are repeated until $c^{n+1}(t)$ is the same as $c^n(t)$. The fitted curve reaches its asymptotic form. This approach allows the fitting to converge and ensures that only the prominences around the curve are counted.

The green bands in Figure~\ref{butterfly_and_driftV_total}(a) highlight the data that are counted for the fitting. Figures~\ref{butterfly_and_driftV_total}(b) and (d), which give the temporal evolution of the monthly average latitude of the high-latitude prominences from 2010 to 2020 in the northern and southern hemispheres, respectively. The derived drift velocity variations in both hemispheres are plotted as red curves in Figures~\ref{butterfly_and_driftV_total}(c) and (e). Note that the positive/negative velocity means that prominences migrate toward the pole/equator in both hemispheres. There are few prominences in high latitudes after 2015 so we do not calculate the drift velocities in this interval. It is seen that the high-latitude prominences migrate toward the polar region with relatively low velocities in the rising phase of solar cycle 24 in both hemispheres. The drift velocities accelerate and reach the peak around mid-2013 in both hemispheres. While the drift velocity decreases after mid-2013 in the southern hemisphere, few prominences are visible in the northern hemisphere.

For the low-latitude prominences, we are unable to obtain a converging latitude range with the same approach, so we fit all the low-latitude prominences directly as in previous works \citep{Li2010, Hao2015}. The temporal evolution of the monthly average latitude of the low-latitude prominences from 2010 to 2019 in the northern and southern hemispheres are shown in Figures~\ref{butterfly_and_driftV_total}(f) and (h), respectively, where the red lines are the corresponding fitted cubic curves. The evolution of the drift velocities in the two hemispheres is displayed in Figures~\ref{butterfly_and_driftV_total}(g) and (i). It is seen that the low-latitude prominences migrate toward the equator with velocities no more than $0.7~\text{m}\ \text{s}^{-1}$ (or $0.005^{\circ}~\text{day}^{-1}$) in the rising phase of solar cycle 24. After reaching the minimum latitude around 2015, the prominences migrate toward high latitudes and the monthly average latitudes of prominences become relatively dispersed. The drift velocities are relative low, approaching $0.7~\text{m}\ \text{s}^{-1}$ in the northern hemisphere, but are accelerated $1.4~\text{m}\ \text{s}^{-1}$ in the southern hemisphere.

	\begin{figure}[ht!]
		\centering
		\includegraphics[width=\linewidth]{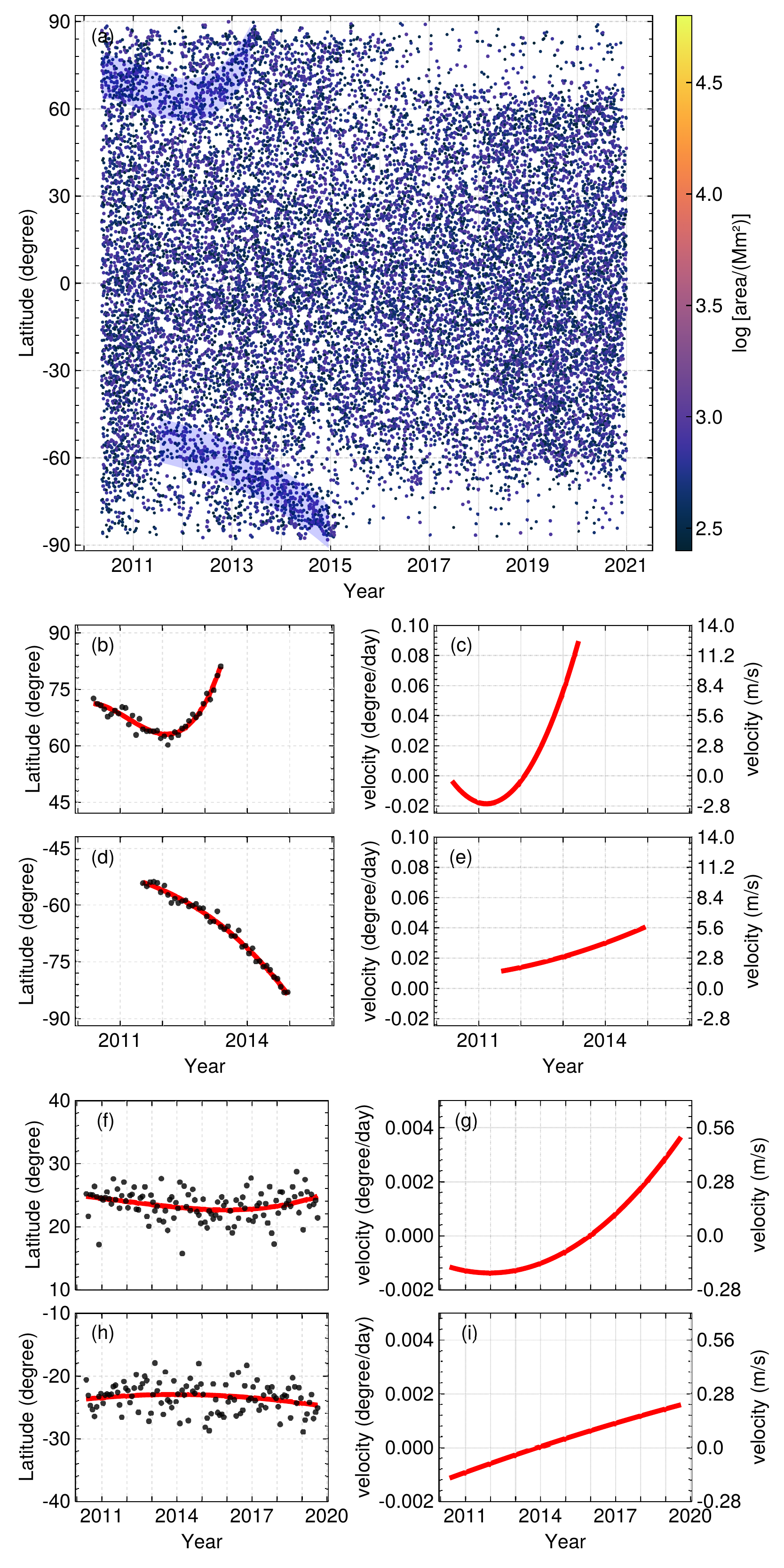}
		\caption{Similar to Figure~\ref{butterfly_and_driftV_total} but for the prominences with areas lower than $1000\ \text{Mm}^2$.}
		\label{butterfly_and_drift_low_small}
	\end{figure}

	\begin{figure}[ht!]
		\centering
		\includegraphics[width=\linewidth]{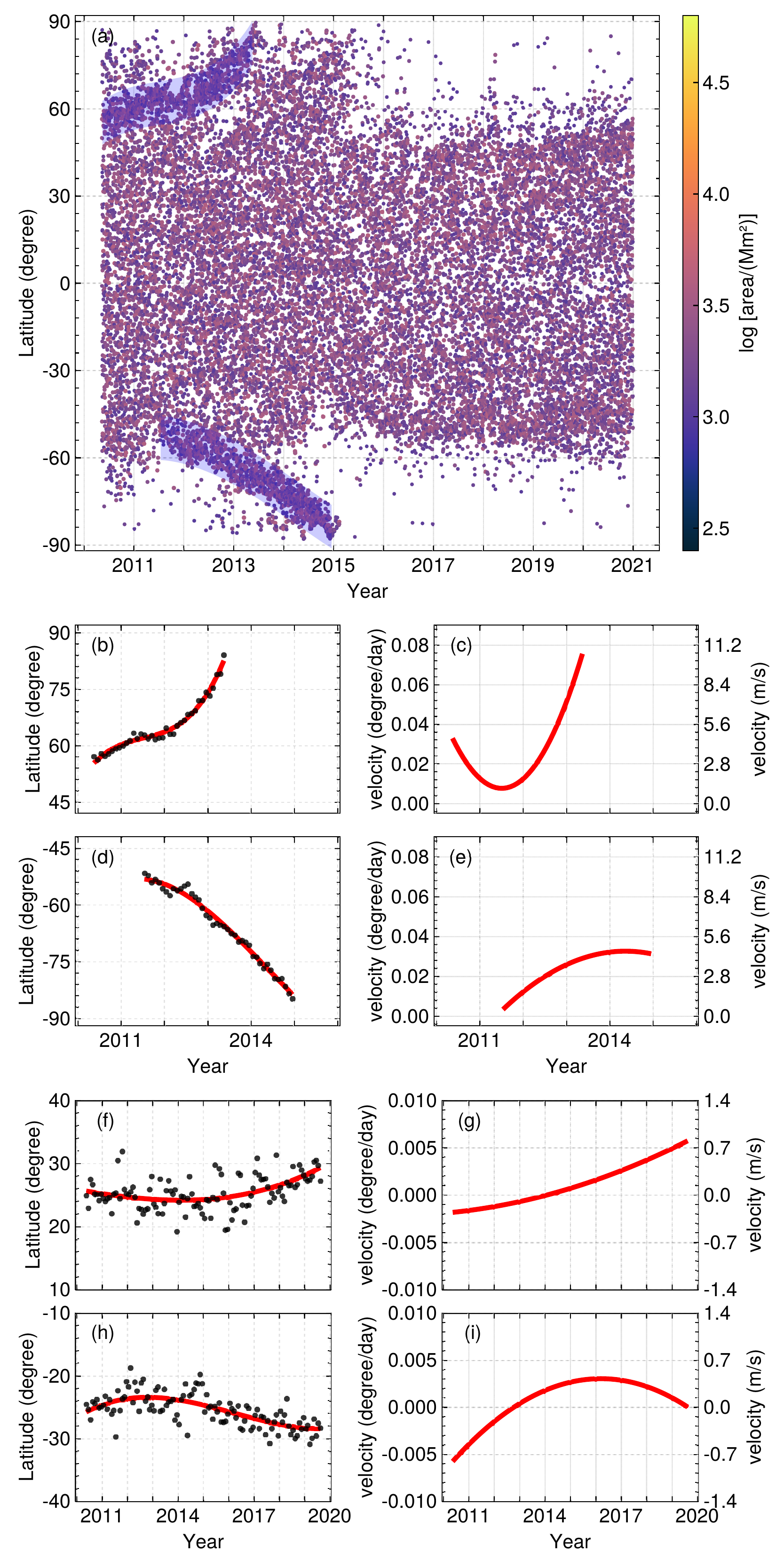}
		\caption{Similar to Figure~\ref{butterfly_and_driftV_total} but for the prominences with areas within $1000--4000\ \text{Mm}^2$.}
		\label{butterfly_and_drift_low_medium}
	\end{figure}

	\begin{figure}[ht!]
		\centering
		\includegraphics[width=\linewidth]{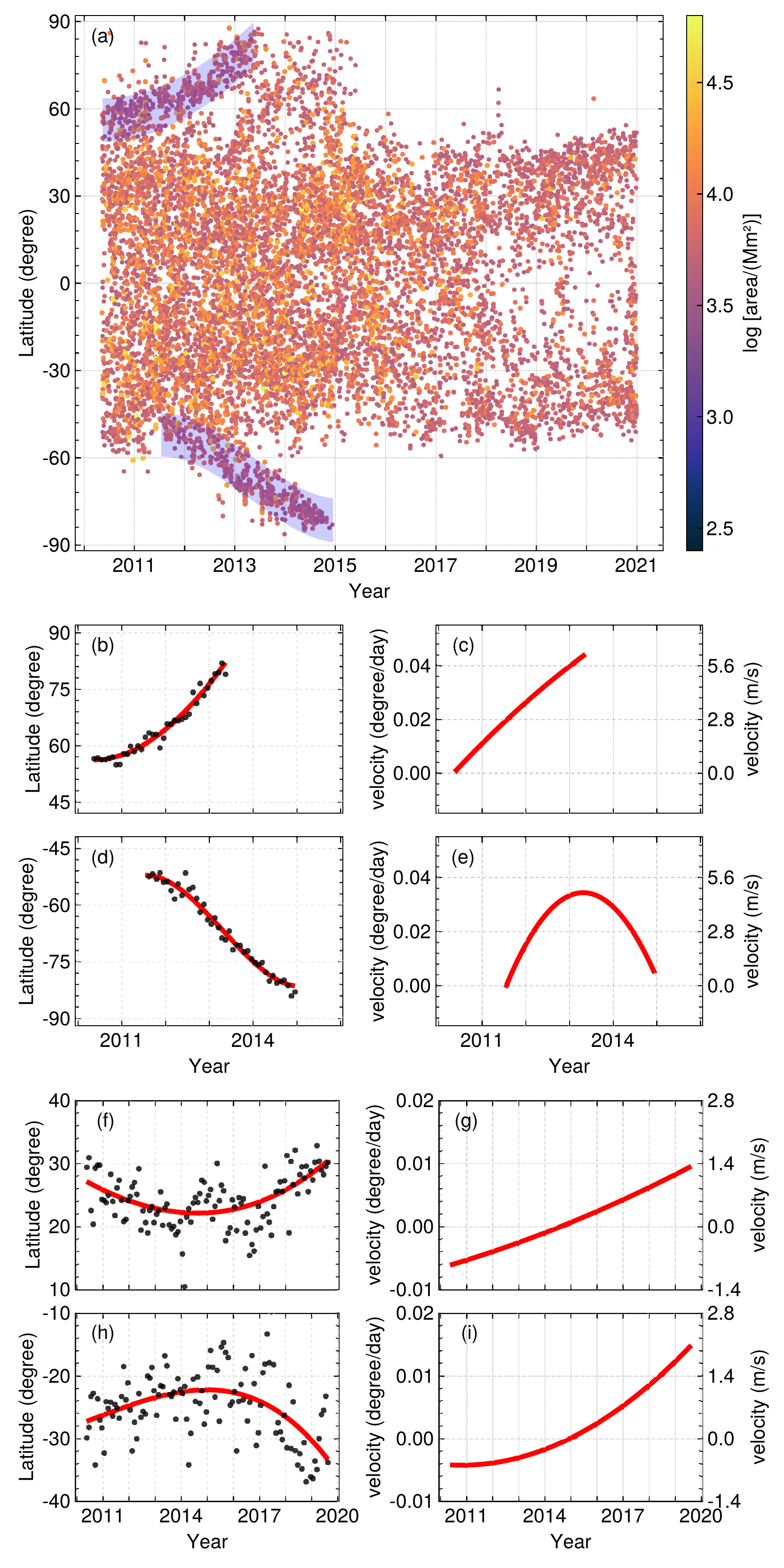}
		\caption{Similar to Figure~\ref{butterfly_and_driftV_total} but for the prominences with areas over $4000\ \text{Mm}^2$.}
		\label{butterfly_and_drift_low_big}
	\end{figure}

We also analyze the drift velocities of the prominences after being divided into three area groups, and the results are shown in Figures~\ref{butterfly_and_drift_low_small}--\ref{butterfly_and_drift_low_big}, respectively. It can be seen that the larger the areas of the prominences are, the clearer the wings of the butterfly diagram are. For the prominences with areas lower than  $1000\ \text{Mm}^2$, considering that they have a much wider latitudinal distribution, we calculate the drift velocities of both high-latitude and low-latitude prominences based on their monthly average latitudes. The high-latitude prominences mainly migrate toward the polar regions. The drift velocities show drastic variation in the northern hemisphere and can be up to $12.4~\text{m}\ \text{s}^{-1}$. On the contrary, the high-latitude prominences in the southern hemisphere migrate toward the polar region with relatively low and steadily increasing velocities from the rising phase to the year around the maximum of solar cycle 24. The low-latitude prominences first migrate toward the equator with a decreasing velocity, then migrate toward the high latitudes with an increasing velocity after reaching the lowest latitude around 22$^\circ$ in around 2015. Although the drift velocity in the northern hemisphere is larger than that in the southern hemisphere, both do not exceed $0.56~\text{m}\ \text{s}^{-1}$. Similar to those with areas lower than  $1000\ \text{Mm}^2$, the drift velocities of the prominences within the area range of  1000--4000 $\text{Mm}^2$ show a clear asymmetry in both hemispheres.
The high-latitude prominences first decelerate and then accelerate in the northern hemisphere while those in the southern hemisphere mainly migrate toward the pole region at an accelerating velocity. The low-latitude prominences migrate toward the equator from 2010 to 2013 and then migrate toward the relatively high latitudes. The migration has two phases in the northern hemisphere, i.e., decelerating in the rising phase and accelerating after the maximum of solar cycle 24. In the southern hemisphere, after migrating toward the equator with a decelerating velocity in the rising phase, the low-latitude prominences migrate toward the relatively high latitudes. Their drift velocities increase first and then decrease in the declining phase of solar cycle 24 after reaching a maximum velocity of around $0.5~\text{m}\ \text{s}^{-1}$ in 2016. Unlike the prominences with areas lower than $4000\ \text{Mm}^2$, the larger prominences show that the drift velocities have similar trends in both hemispheres.

	The high-latitude prominences with areas larger than $4000\ \text{Mm}^2$ migrate toward the polar regions. Their drift velocities increase in the rising phase, reaching a maximum of around $5.6~\text{m}\ \text{s}^{-1}$ around the maximum of solar cycle 24 in later 2013. Due to the difference in the time intervals of fitting, the southern hemisphere shows a deceleration of the velocities towards the polar region afterward. The low-latitude prominence with areas larger than $4000\ \text{Mm}^2$ migrate toward the equator at a decelerating velocity in the rising phase, and then turn to migrate toward relatively high latitudes at an accelerating velocity after solar maximum. Besides, for low-latitude prominences, the drift velocities of the larger prominences are larger than those of the smaller prominences, especially in the declining phase of solar cycle 24, and the maximum velocity can be up to $2.1~\text{m}\ \text{s}^{-1}$. It is seen from the quantitatively calculated drift velocities in panels (f--i) of Figures~\ref{butterfly_and_drift_low_small}--\ref{butterfly_and_drift_low_big}.

	\subsubsection{Morphology of Prominences}

	\begin{figure}[ht!]
		\centering
		 \includegraphics[width=\linewidth]{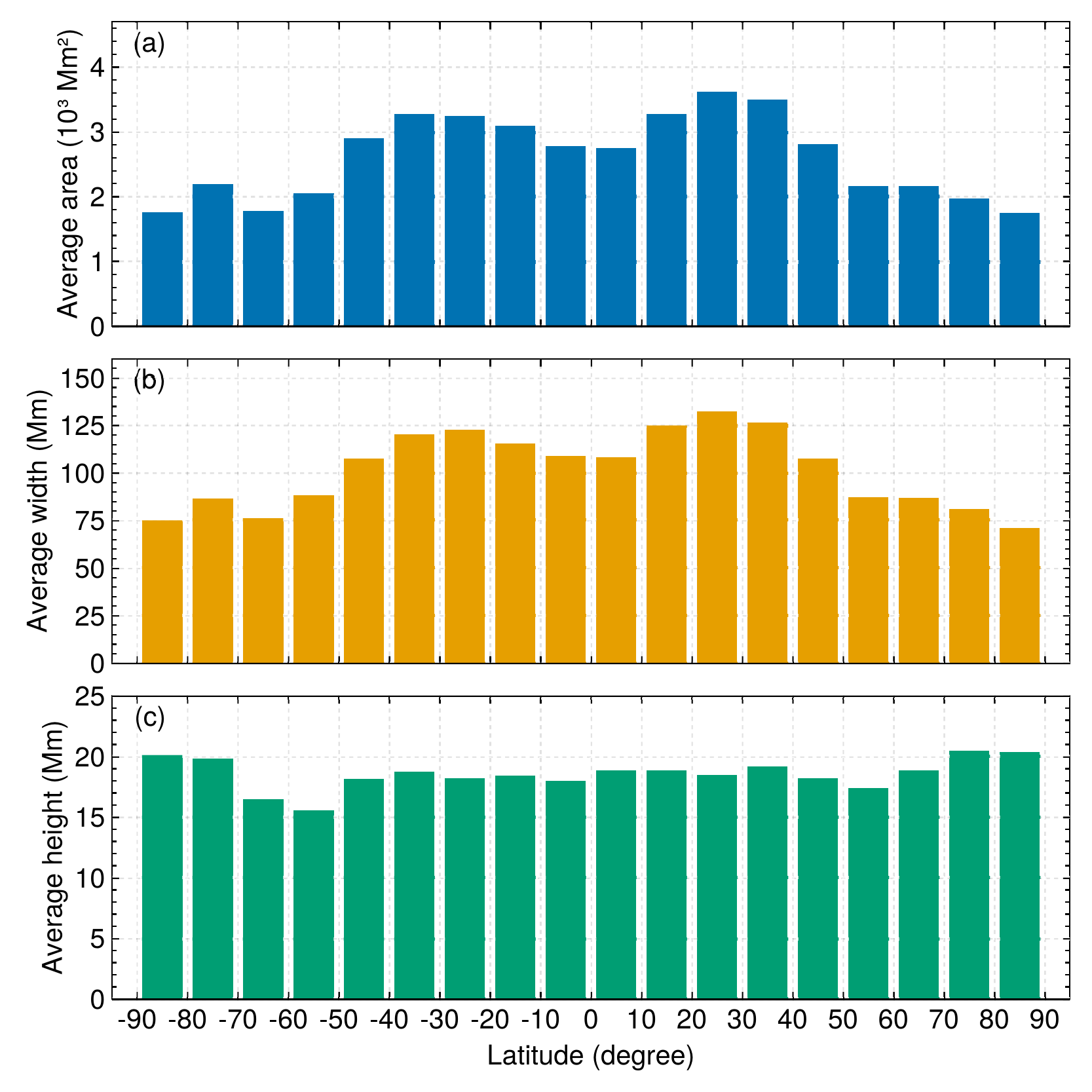}
		\caption{Distributions of the prominences features with respect to latitude. Panels (a), (b), and (c) show the average area, width, and height of the prominences in different latitude bands, respectively.}
		\label{p_average_area_width_height_latitude}
	\end{figure}

	\begin{figure}[ht!]
		\centering
		\includegraphics[width=\linewidth]{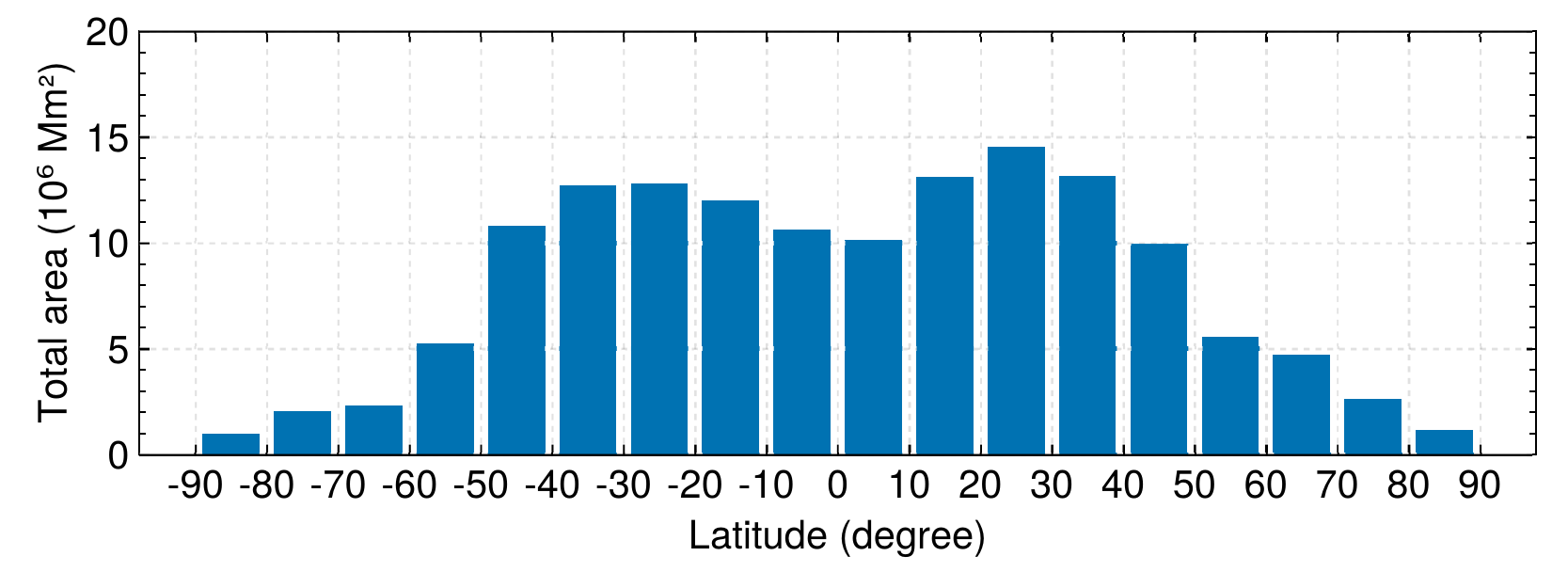}
		\caption{Similar to Figure~\ref{p_average_area_width_height_latitude} but for the cumulative area.}
		\label{p_total_area_latitude}
	\end{figure}

	\begin{figure}[ht!]
		\centering
		\includegraphics[width=\linewidth]{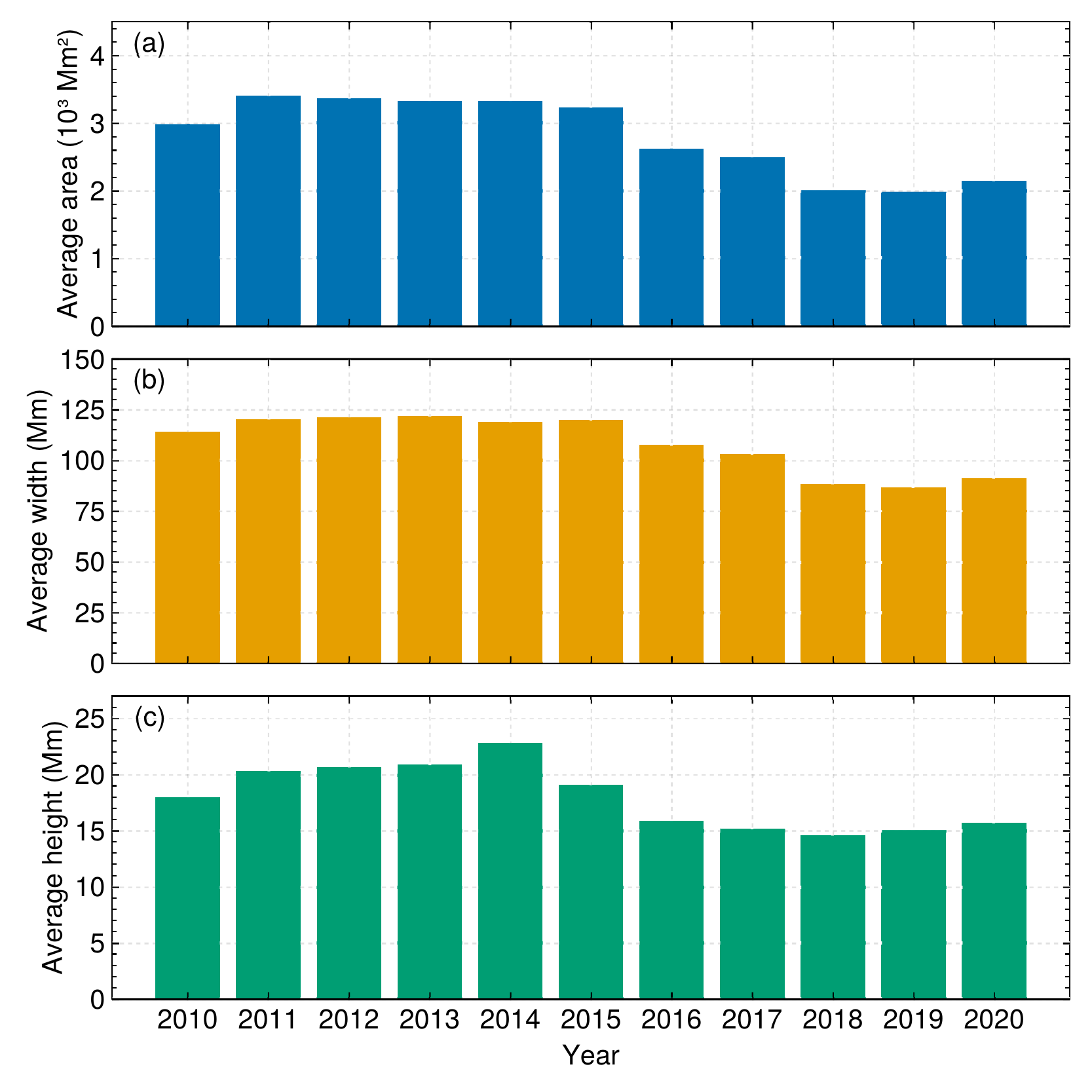}
		\caption{Distributions of the prominences features with respect to years. Panels (a), (b), and (c) show the average area, width, and height of the prominences in different years, respectively.}
		\label{p_average_area_width_height_year}
	\end{figure}

	\begin{figure}[ht!]
		\centering
		\includegraphics[width=\linewidth]{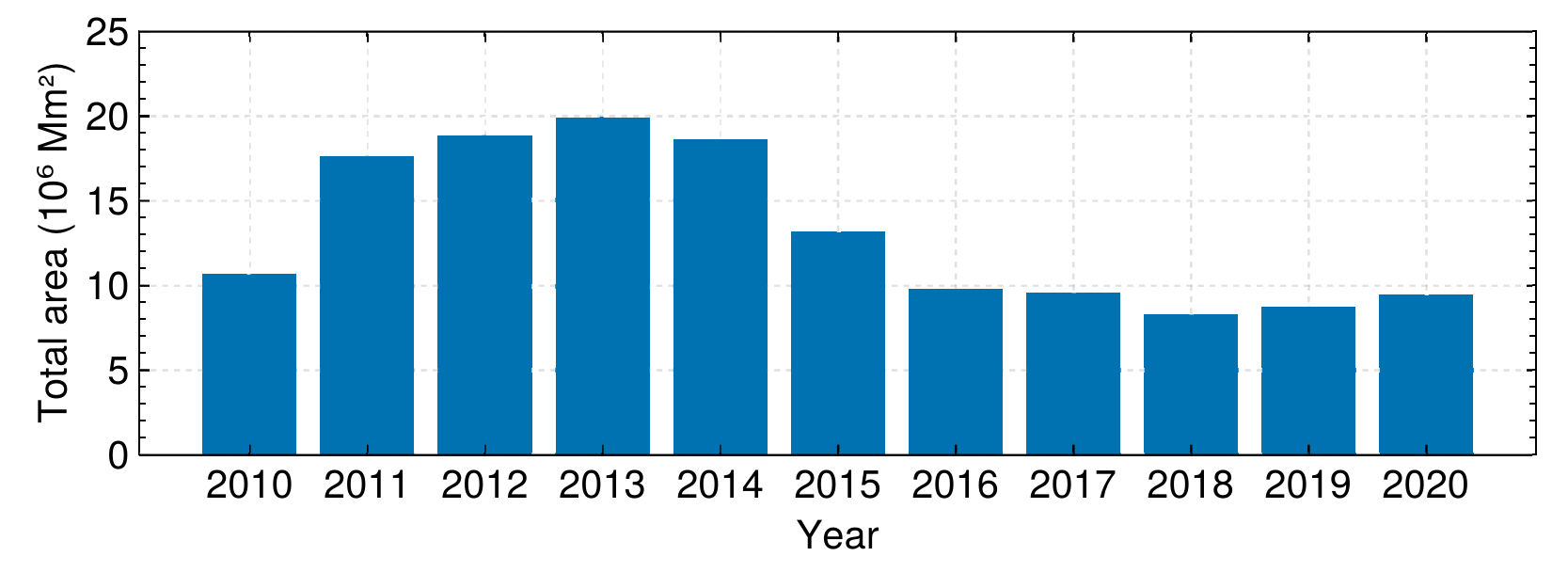}
		\caption{Similar to Figure~\ref{p_average_area_width_height_year} but for the cumulative area. }
		\label{p_total_area_year}
	\end{figure}

	In this subsection, we analyze the prominence area, height, and width variations with respect to latitude bands and time. We detect prominences from the transformed images, where the horizontal coordinate represents the position angle measured clockwise from the south pole and the vertical coordinate represents the heliocentric distance, i.e., the $x$-axis of the rectangle is along the solar limb and the $y$-axis is along the radial direction. In this way, the thickness of a prominence equals the top-height minus the bottom-height of the prominence, and the angular width refers to the angular span of the prominence along the solar limb. We also calculate the geometric center height of the prominence as its characteristic height. It is noted that the area, height, and width of the prominence here are the projection values, and a real prominence is a three-dimensional structure.

	The histograms in Figure~\ref{p_average_area_width_height_latitude} show that the average areas and widths are roughly bimodally distributed. The largest average area is about $3600~\text{Mm}^2$ and the largest average width is $130\ \text{Mm}$, which appears in the latitude band of $20^{\circ}$ to $30^{\circ}$ in both hemispheres, though the largest average area in the southern hemisphere appears in the latitude band of $30^{\circ}$ to $40^{\circ}$. It indicates that the prominences are relatively narrow and small in high latitudes and around the equator. The latitudinal distribution of the average height shows weak variation in the latitudes lower than $50^{\circ}$ with typical height being $18\ \text{Mm}$. However, the average height of the prominences in the latitude bands higher than $50^{\circ}$ gradually increases with the increasing latitude and peaks around the polar regions in high latitude bands higher than $50^{\circ}$. It indicates that the prominences in high latitudes are higher. The cumulative area of the prominences in different latitude bands also presents a bimodal distribution as shown in Figure~\ref{p_total_area_latitude}. The largest cumulated areas appear in the latitude band of $20^{\circ}$ to $30^{\circ}$  in both hemispheres. It indicates that the prominences are more frequently formed in the latitude band of  $20^{\circ}$ to $30^{\circ}$. However, the cumulative area decreases more rapidly with increasing latitude than the average area.

	Figure~\ref{p_average_area_width_height_year} shows the average area, width, and height with respect to years in panels (a--c). The average area and width show a decreasing trend from the beginning to the end of solar cycle 24 in general. The maximum average area is $3408~\text{Mm}^2$, which appears in 2011. The maximum average width appears in 2013, which is $120~\text{Mm}$. Actually, there is no significant variation in the average area and width from 2011 to 2015. The average height gradually increases from 2010 to 2014 with a peak value of $23~\text{Mm}$ in 2014. After that, it decreases in the following two years and remains around $15~\text{Mm}$. It indicates that the prominences become lower in the declining phase of solar cycle 24. The cumulated areas in different years are plotted in Figure~\ref{p_total_area_year}. The largest cumulated area appears in 2013, which is consistent with the results of prominence number in Figure~\ref{p_number_year}.

	\begin{figure}[ht!]
		\centering
		\includegraphics[width=\linewidth]{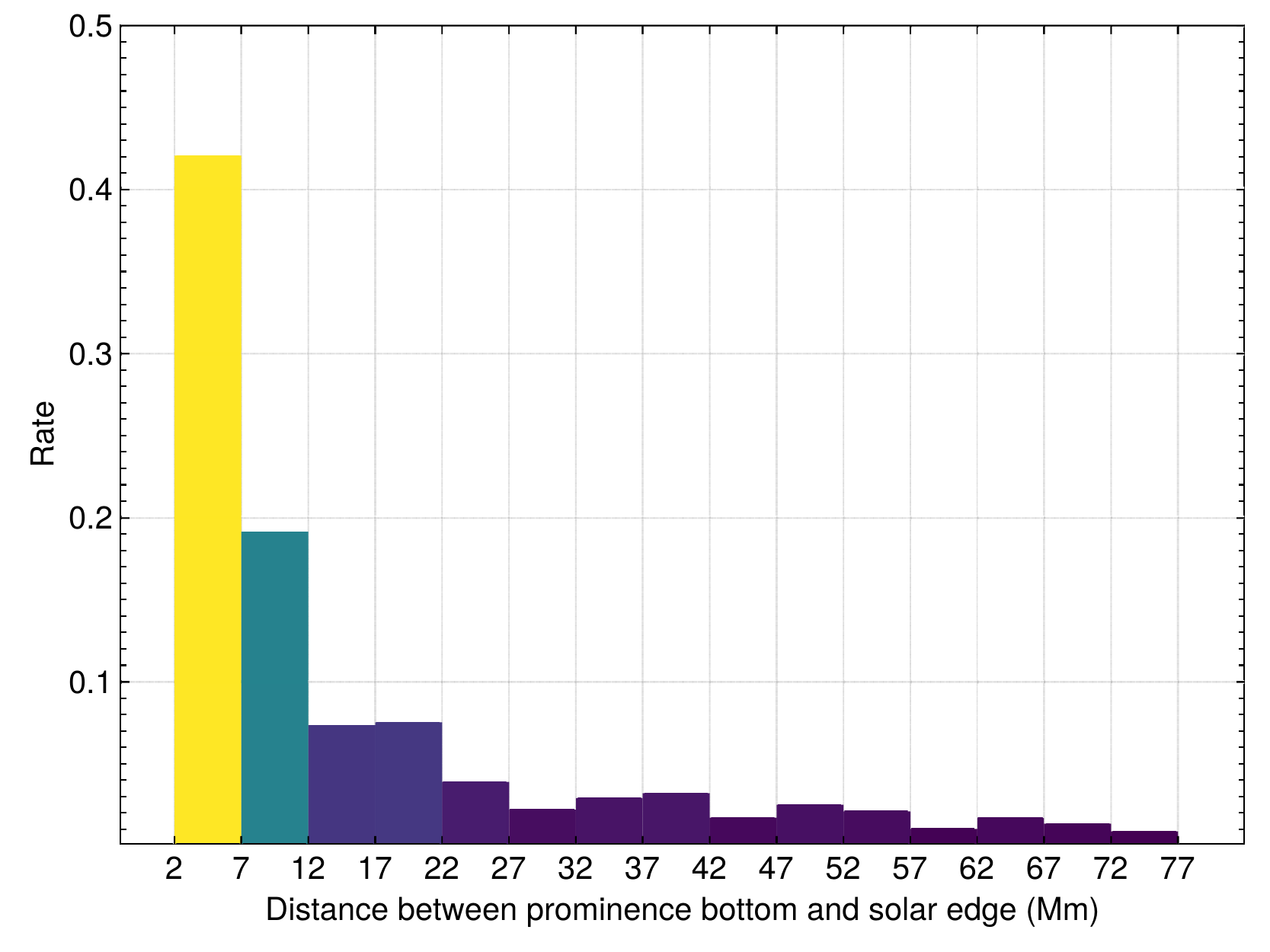}
		\caption{Variations of the distance between the bottom of prominence and solar limb.}
		\label{distance_from_limb}
	\end{figure}

	Prominences are dense ``clouds" suspended in the corona frequently with several legs connected to the solar limb \citep{Parenti2014, Vial2015}. However, sometimes they are not connected to the solar limb. We calculate the distance between the bottom-height of the detected prominences and the solar limb and draw a histogram of such a distance in Figure~\ref{distance_from_limb}. Here we define that if the distance between the bottom of a prominence and the solar limb is more than $2~\text{Mm}$, the prominence is considered to be not connected to the solar limb. The $2~\text{Mm}$ criterion is determined by considering the detection error. It is found that $10\%$ of the prominences are totally detached from the solar limb.

	\begin{figure}[ht!]
		\centering
		\includegraphics[width=\linewidth]{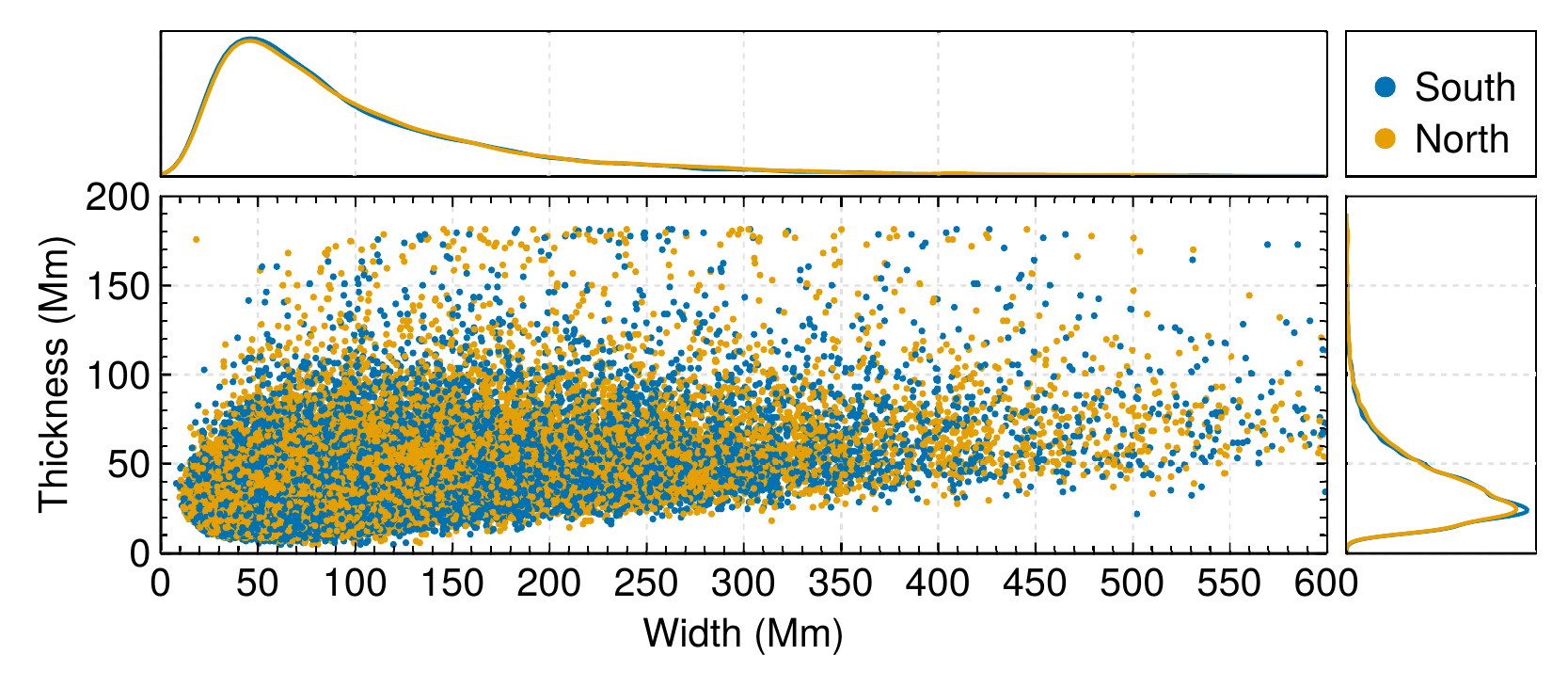}
		\caption{Relationship between the thickness and width of the prominences. Each dot shows a single prominence. The blue and orange dots represent the observations in the southern and northern hemispheres, respectively. The right panel is the percentage profile with respect to the prominence thickness. The top panel is the percentage profile with respect to the prominence width. The blue and orange lines represent the results in the southern and northern hemispheres, respectively.}
		\label{p_height_width}
	\end{figure}

	\begin{figure}[ht!]
		\centering
		\includegraphics[width=\linewidth]{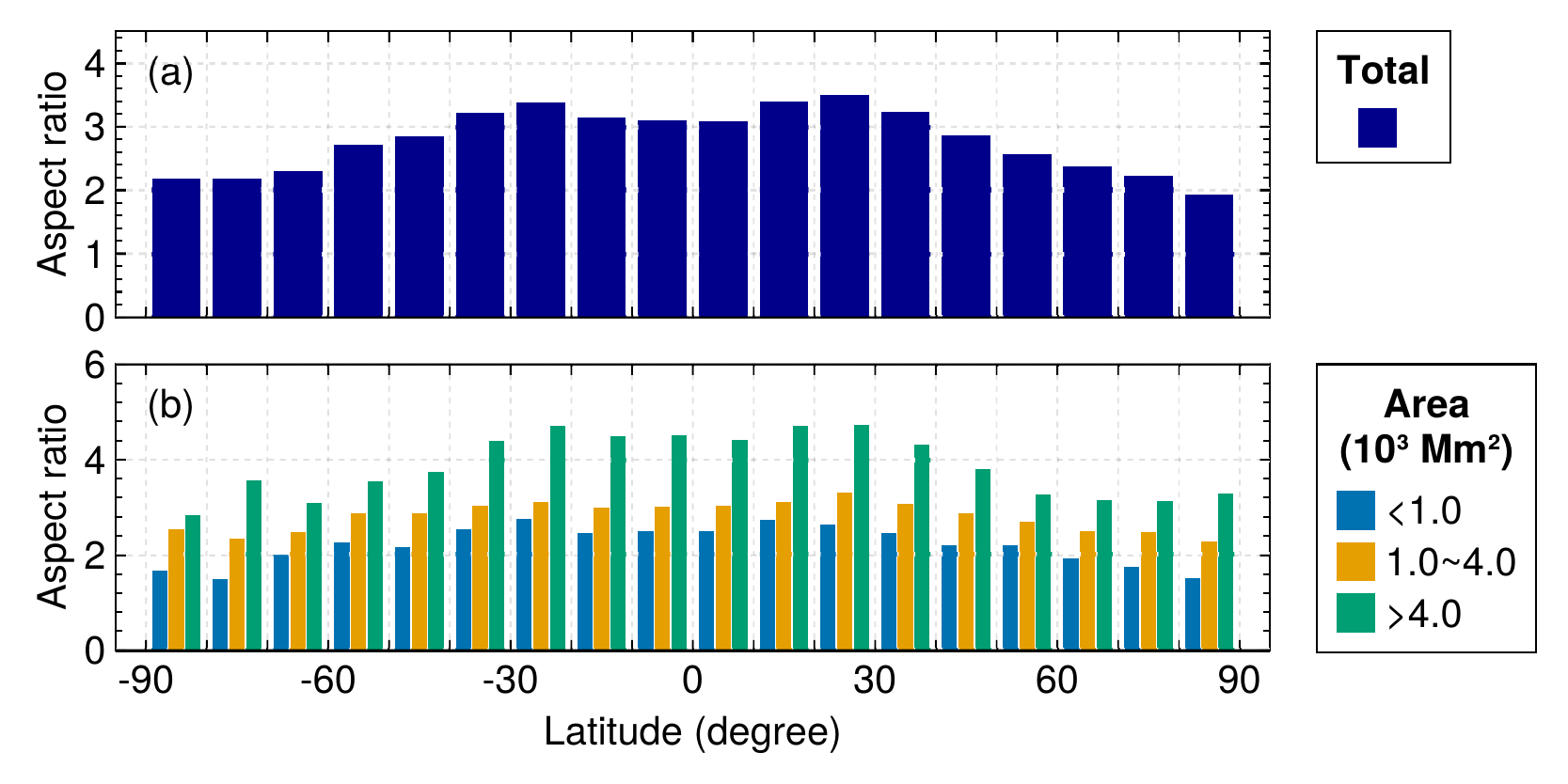}
		\caption{Variation of the aspect ratio of the prominences with respect to latitude.}
		\label{p_wh_ratio_latitude}
	\end{figure}

	\begin{figure}[ht!]
		\centering
		\includegraphics[width=\linewidth]{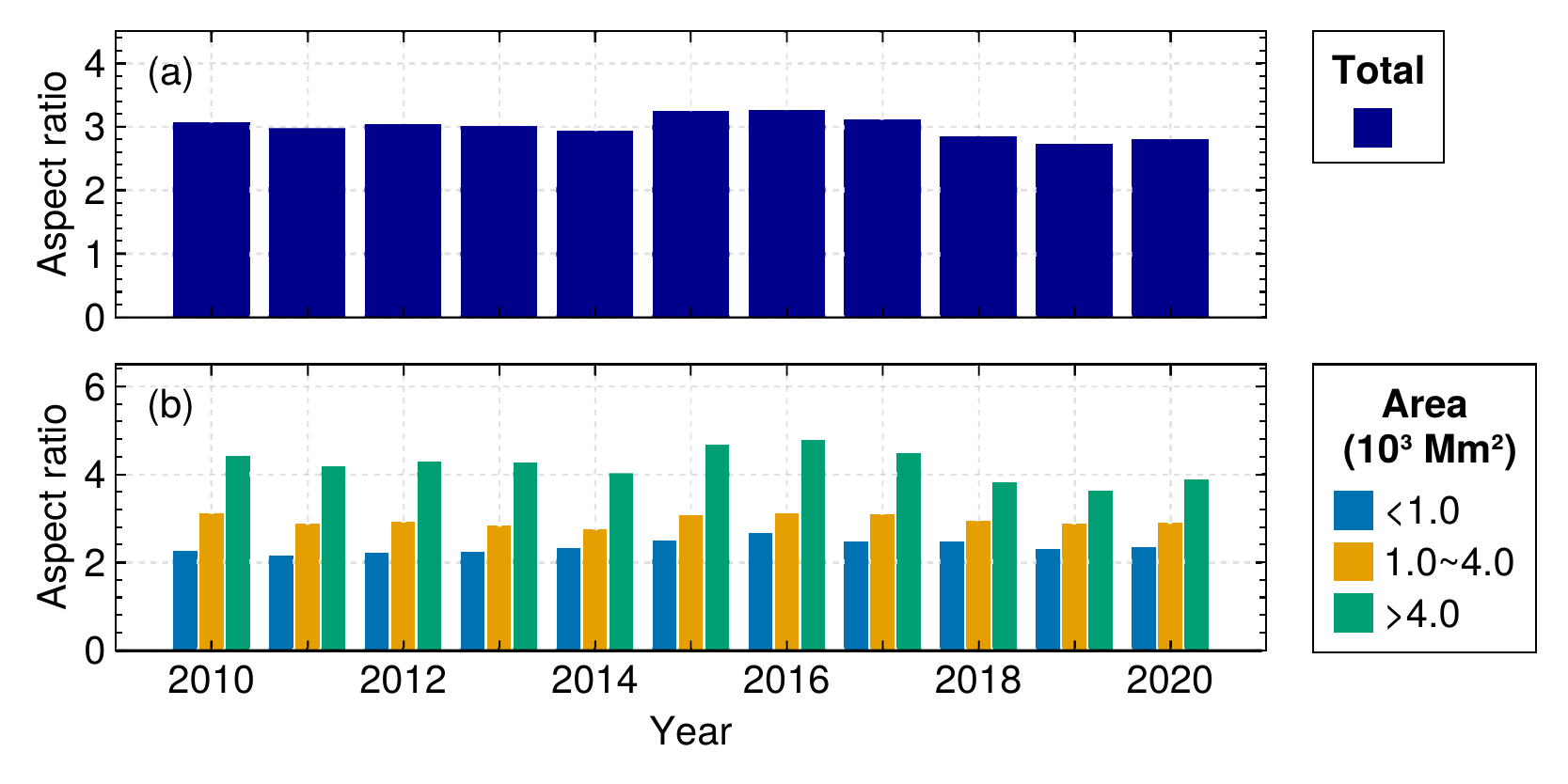}
		\caption{Variation of the aspect ratio of the prominences with respect to years.}
		\label{p_wh_ratio_year}
	\end{figure}

	We also analyze the relationship between the thickness and the width of prominence, which is shown in Figure~\ref{p_height_width}. Most prominences have widths below $400~\text{Mm}$ and their thicknesses are below $100~\text{Mm}$. The right and top panels display the percentage of prominences with different thicknesses and widths in the northern (orange line) and southern (blue line) hemispheres, respectively. The percentage profiles of prominences with respect to the thickness in the northern and southern hemispheres are almost identical. The thickness distribution peaks at around $25~\text{Mm}$ and the width distribution peaks at around $48~\text{Mm}$.

	The ratio of the width to the thickness of a prominence can be used as an indicator of the shape of the prominence, which we defined as the aspect ratio in our analysis. Figures~\ref{p_wh_ratio_latitude} and \ref{p_wh_ratio_year} show the variation of the aspect ratio of the prominences in different latitude bands and time. The latitudinal distribution of the aspect ratio is also bimodal. Prominences appear to be wider and thinner in the latitude band of  $20^{\circ}$ to $30^{\circ}$ and the reverse is true in the high latitude bands in both hemispheres. This conclusion also holds for the prominences with different area ranges. However, there are small peaks in high latitudes for the prominences with areas lower than $1000\ \text{Mm}^2$. Figure~\ref{p_wh_ratio_year}(a) shows that the aspect ratio of the prominence remains quite stable throughout solar cycle 24 with a small peak around 2016. The prominences with different area ranges also show little variation in the aspect ratio as illustrated in Figure~\ref{p_wh_ratio_year}(b).

\subsubsection{N-S Asymmetry of Prominences}

The N-S asymmetry is a typical characteristic of various solar features, such as the sunspot number and area, filament and flare occurrences, and so on, which are unidentical over the northern and southern hemispheres \citep{Li2010, Hao2015, Chowdhury2019, Roy2020, Veronig2021, Chandra2022}. The N-S asymmetry is observed in different phases of a solar cycle \citep{Hao2015, Chowdhury2019, Veronig2021, Chandra2022}.

	We also investigate the N-S asymmetry of the prominence number and the cumulative prominence area. The N-S asymmetry index is defined as
	\begin{equation}
		A_n=\frac{N_n-N_s}{N_n+N_s},
	\end{equation}
	\begin{equation}
		A_c=\frac{C_n-C_s}{C_n+C_s},
	\end{equation}
	where $N$ and $C$ represent the prominence number and the cumulative area, while $n$ and $s$ represent the northern and southern hemispheres, respectively.  $A_n$ is the N-S asymmetry index for the prominence number and $A_c$  is the N-S asymmetry index for the normalized cumulative area, respectively. If the sign of the N-S asymmetry index is positive, it means that the northern hemisphere is the dominant hemisphere, and vice versa.

\begin{figure}[ht!]
		\centering
		\includegraphics[width=\linewidth]{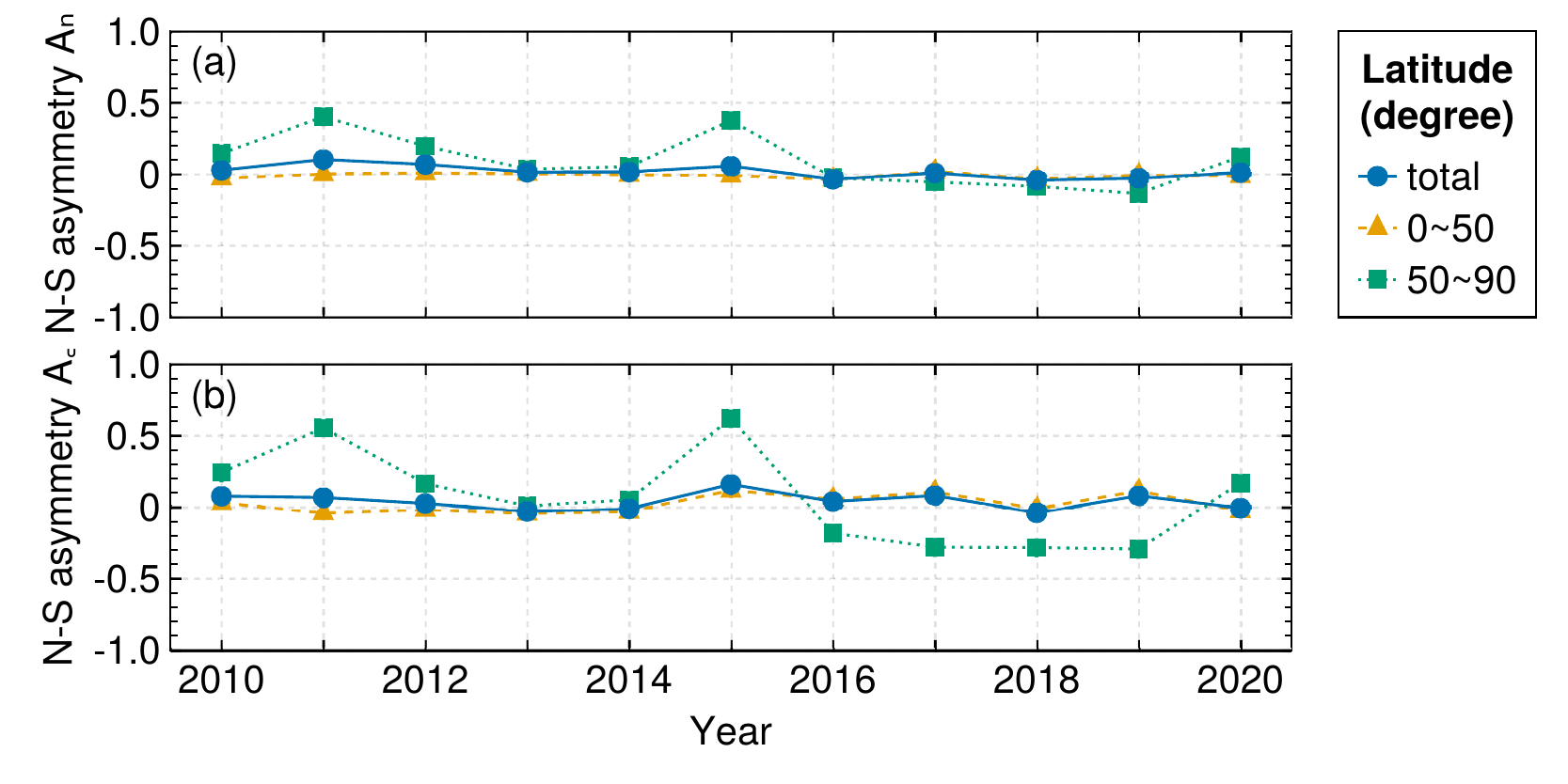}
		\caption{Yearly asymmetries of the prominence features with respect to latitude from 2010 to 2020. (a) Yearly asymmetry indices of the prominence numbers in different latitude bands. (b) Yearly asymmetry indices of the cumulative prominence areas in different latitude bands. The solid lines marked with circles, dash lines marked with triangles, and dot lines marked with squares represent all prominences, low and high-latitude prominences, respectively.}
		\label{p_nsa}
	\end{figure}

	The yearly asymmetries of the prominence number and cumulative area in different latitude bands are plotted in Figure~\ref{p_nsa}.  The figure indicates that the two hemispheres are almost identical in terms of either total prominence number or total cumulative area. However,  the high-latitude prominences show much stronger N-S asymmetry, e.g., in terms of the prominence number, the northern hemisphere is dominant in $\sim$2011 and $\sim$2015, and in terms of the prominence cumulative area, the northern hemisphere is dominant in $\sim$2011 and $\sim$2015, but the southern hemisphere is dominant during 2016--2019. The indices for the low-latitude prominences are similar to those of the total prominences since most of the prominences are within the latitude band of $0^{\circ}$ to $50^{\circ}$. It is noticed that the asymmetry indices of cumulative prominence area have a similar trend to those of prominence number but have larger values.

	\subsection{Active Regions}

	\subsubsection{Distributions of the Number of Active Regions}

	\begin{figure}[ht!]
		\centering
		\includegraphics[width=\linewidth]{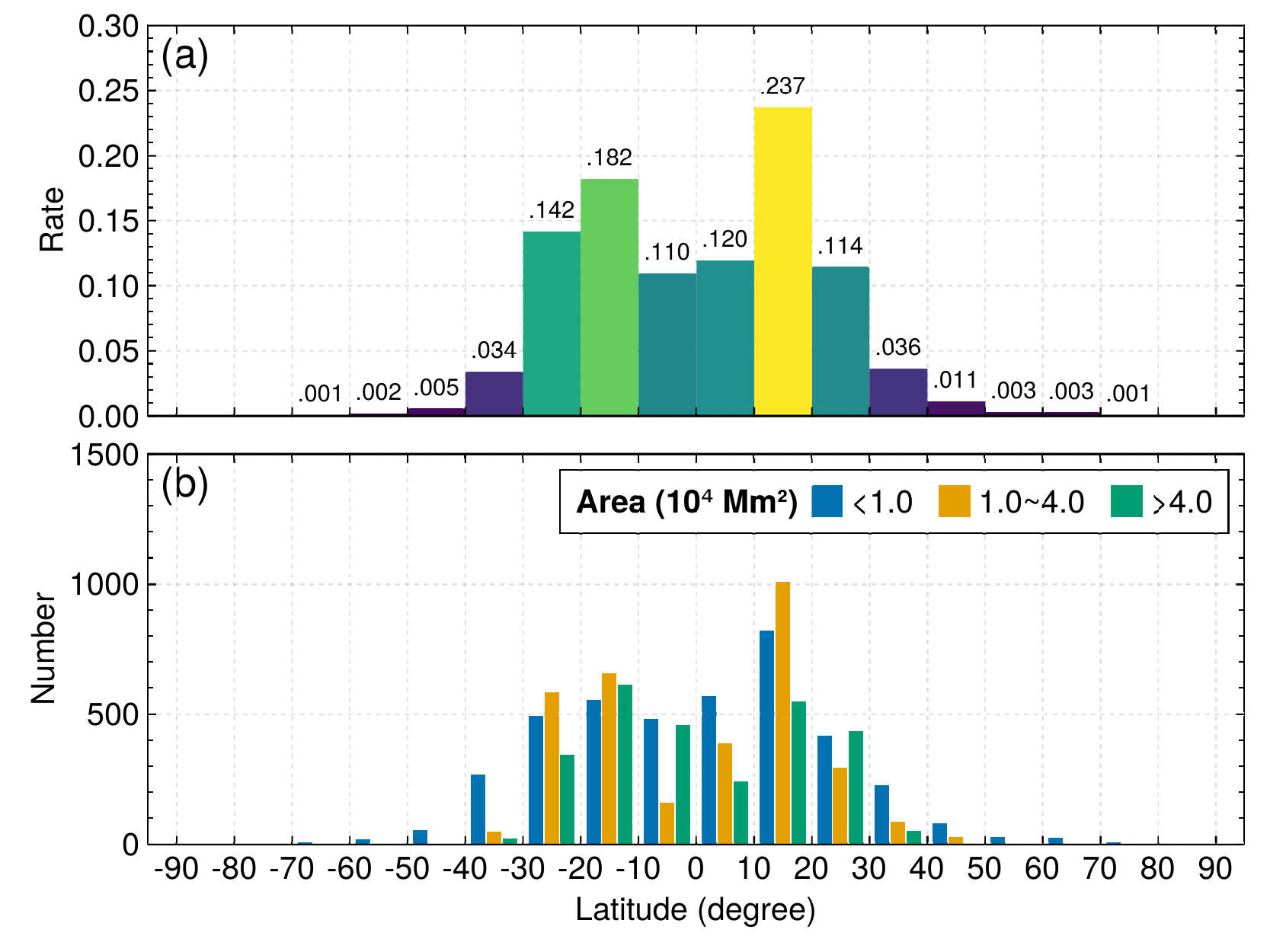}
		\caption{Distribution of the AR number with respect to latitude. (a) The normalized AR number in the different latitude bands. (b) Distribution of the AR number within three area groups in different latitude bands.}
		\label{ar_number_latitude}
	\end{figure}

	\begin{figure}[ht!]
		\centering
		\includegraphics[width=\linewidth]{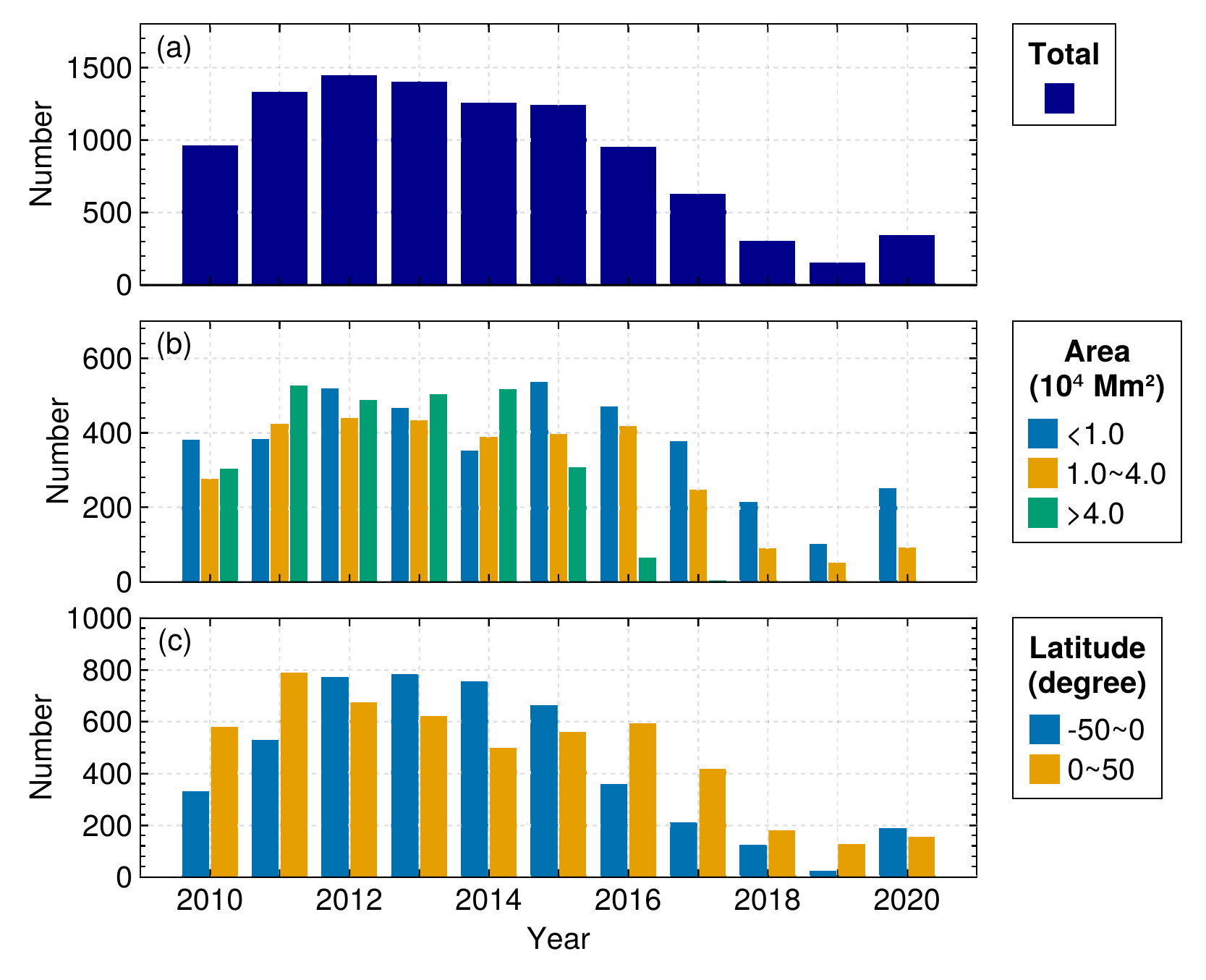}
		\caption{Distribution of the AR number from 2010 to 2020. (a) The cumulative AR number in each year. Panels (b) and (c) are similar to panel (a) but for the number of ARs which are grouped by areas and latitude bands, respectively.}
		\label{ar_number_year}
	\end{figure}

	We have detected $10,018$ ARs from 2010 May 13 to 2020 December 31. The ARs in our study refer to the regions that have enhanced emissions but with differences in morphology and appearance compared to prominences, e.g., the former are less brighter and compacter than prominences. Since the detection is taken in the off-limb regions, the emissions are slightly brighter than the surrounding quiet corona. As our detection method is very sensitive, these regions would also be identified and recognized in our study.

	As done for prominences, we draw the latitudinal and yearly distributions of ARs in Figures~\ref{ar_number_latitude} and ~\ref{ar_number_year}. The number of ARs has peaks in the latitude band of $10^{\circ}$ to $20^{\circ}$ with a bimodal distribution. About $99.1\%$ of the detected ARs have latitudes lower than $50^{\circ}$. The detected ARs are also divided into three groups according to their areas, i.e., $<$$1\times 10^{4}\ \text{Mm}^2$, $1\times 10^{4}$--$4\times 10^{4}\ \text{Mm}^2$ and $>$$4\times 10^{4}\ \text{Mm}^2$, whose proportions are about $10:8:7$. Figure~\ref{ar_number_latitude}(b) shows the latitudinal distributions of the number of ARs in the three area ranges. Their distributions are similar to the total number distribution in Figure~\ref{ar_number_latitude}(a), i.e., a bimodal distribution with peaks in the latitude band of $10^{\circ}$ to $20^{\circ}$. It is noted that the histograms in the northern and southern hemispheres are not symmetric, especially for the moderate ARs with areas within $1\times10^{4}$--$4\times 10^{4}\ \text{Mm}^2$,  which appear more predominantly in latitude band of $10^{\circ}$ to $20^{\circ}$ of the northern hemisphere. There are few ARs with areas over $1 \times 10^{4}\ \text{Mm}^2$ appearing in the latitudes higher than $50^{\circ}$. It indicates that only small ARs with relatively weak emissions appear in high latitudes.

	Figure~\ref{ar_number_year} shows the yearly distributions of the number of ARs within solar cycle 24. As shown in Figure~\ref{ar_number_year}(a), the AR number increases from the year 2010 and reaches the peak in the year 2012, then decreases in the declining phase. It increases again from the year 2019 to 2020, which indicates the beginning of new solar cycle 25. From Figure~\ref{ar_number_year}(b), we can see the ARs within the three area ranges have similar variation trends with two peaks. The only slight difference is the year when the peaks occur: the ARs with areas lower than $<$$1 \times 10^{4}\ \text{Mm}^2$ have peaks in 2012 and 2015; the ARs with areas within $1 \times 10^{4}$--$4 \times 10^{4}\ \text{Mm}^2$ have peaks in 2012 and 2016 and the ARs with areas over $4 \times 10^{4}\ \text{Mm}^2$ have peaks in 2011 and 2014. From the year 2017 to 2020 the number of the moderate and large ARs is extremely small. These results indicate that ARs are relatively small in the declining phase of solar cycle 24. The yearly distributions of the AR number in the low latitude bands are plotted in Figure~\ref{ar_number_year}(c), which shows that the histograms in the northern and southern hemispheres are asymmetric. The peak number in the northern hemisphere appears in the year 2011 and that in the southern hemisphere appears in the year 2013 for the ARs lower than $50^{\circ}$. Since almost all ARs are located lower than $50^{\circ}$, we do not show the results of ARs higher than $50^{\circ}$.

\subsubsection{Butterfly Diagram of Active Regions}

\begin{figure*}[ht!]
		\centering
		\includegraphics[width=\linewidth]{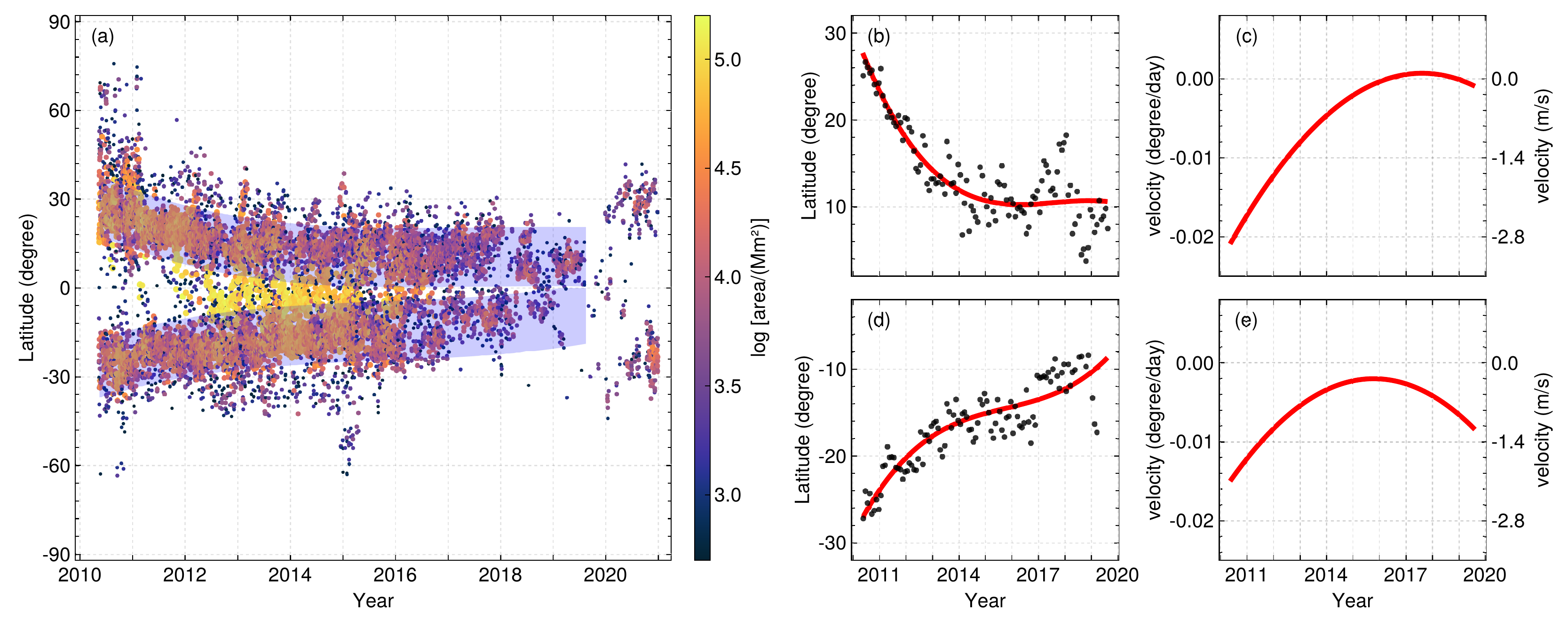}
		\caption{Butterfly diagram and latitudinal migration of ARs. (a) Butterfly diagram of ARs from 2010 to 2020. Each dot represents a single AR. The colorbar shows the logarithm of AR areas.  The blue bands show the data used to calculate the drift velocities of the AR. (b) Temporal evolution of the monthly average latitude distributions of the AR from 2010 to 2020 in the northern hemisphere. Cubic polynomial fittings as shown as the red lines. (c) Drift velocity variations of the monthly average latitude distributions of the AR from 2010 to 2020 in the northern hemisphere. Panels (d) and (e) are similar to (b) and (c) but for the ARs in the southern hemisphere.}
		\label{butterfly_and_driftV_ar}
	\end{figure*}

	The butterfly diagram of ARs is plotted in Figure~\ref{butterfly_and_driftV_ar}. Each dot in the figure represents a single AR, where the latitude is taken from its geometric center. It is seen that ARs are mainly distributed in the latitude bands lower than $40^{\circ}$ in both hemispheres and migrate toward the equator in each solar cycle. We can also find that some ARs are located in high latitudes over $60^{\circ}$, though there are only a few. Larger ARs are mainly distributed in the low latitudes around the equator, which are shown as the yellow dots in Figure~\ref{butterfly_and_driftV_ar}(a). The ARs appear in relatively high latitudes after 2019, marking the beginning of solar cycle 25.

	We calculate the monthly average latitudes of the ARs in both hemispheres to analyze the latitudinal migration and drift velocities quantitatively. The fitting method is the same as that for the high-latitude prominences, but the $\delta$ is set to $10^{\circ}$. The blue bands cover the data that are taken into account for the fitting. Figures~\ref{butterfly_and_driftV_ar}(b) and (d) display the temporal evolutions of the monthly average latitude of the ARs from 2010 to 2019 in the northern and southern hemispheres, respectively. The derived drift velocities in both hemispheres are plotted as red curves in Figures~\ref{butterfly_and_driftV_ar}(c) and (e). It is seen that ARs migrate toward the equator throughout the whole solar cycle 24 with relatively low velocities of 0.5--2.8~$\text{m}\ \text{s}^{-1}$ (or $0.003^{\circ}$--$0.02^{\circ}~\text{day}^{-1}$). In the rising phase, the drift velocity in the northern hemisphere is higher than that in the southern hemisphere from the year 2010 to 2015. After the maximum around late 2015, the velocity increases in the southern hemisphere and the migration in the southern hemisphere is faster than that in the northern hemisphere.

\subsubsection{Morphology of Active Regions}

	\begin{figure}[ht!]
		\centering
		 \includegraphics[width=\linewidth]{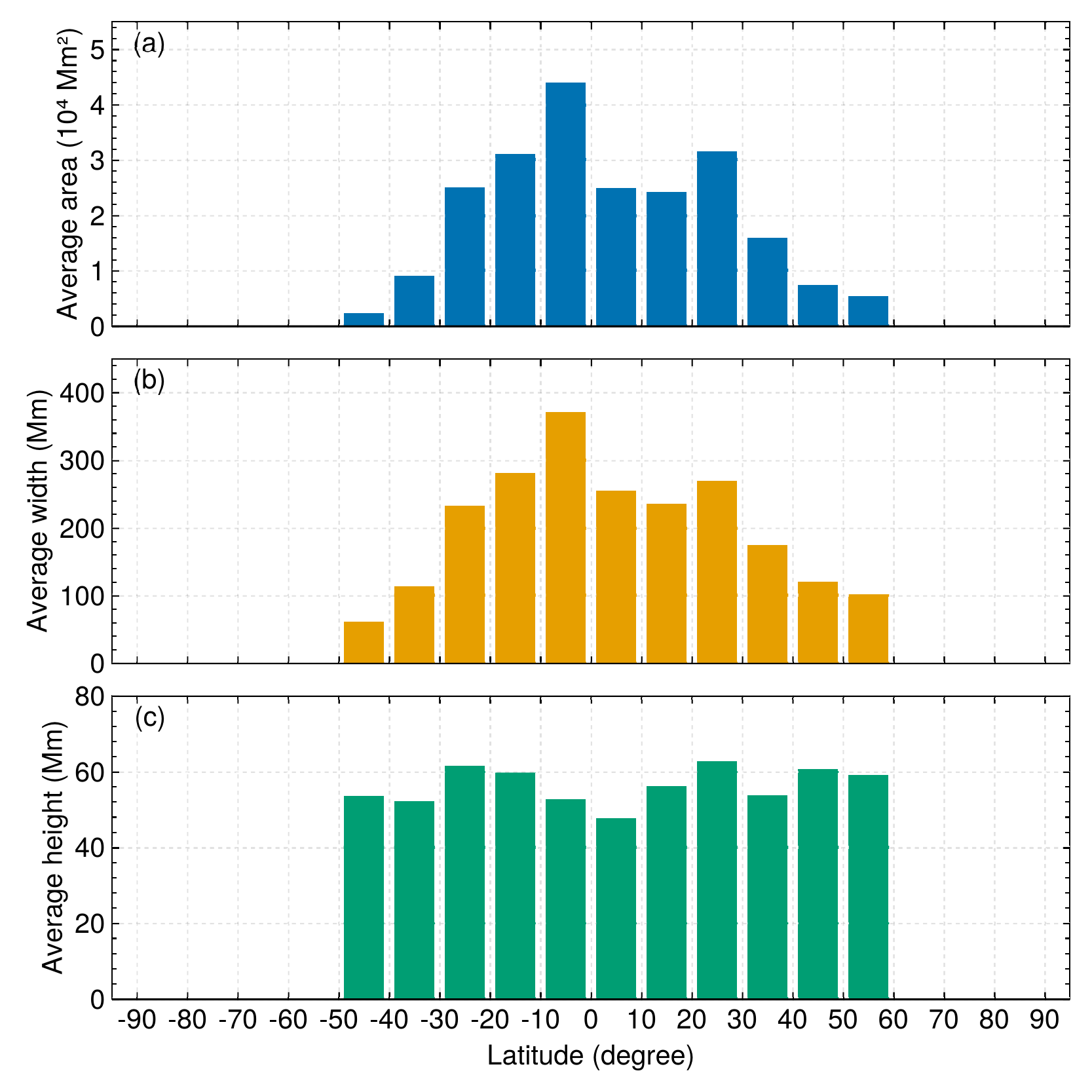}
		\caption{Distributions of the AR features with respect to latitude. Panels (a), (b), and (c) show the average area, width, and height of the ARs in different latitude bands, respectively.}
		\label{ar_average_area_width_latitude}
	\end{figure}

	\begin{figure}[ht!]
		\centering
		\includegraphics[width=\linewidth]{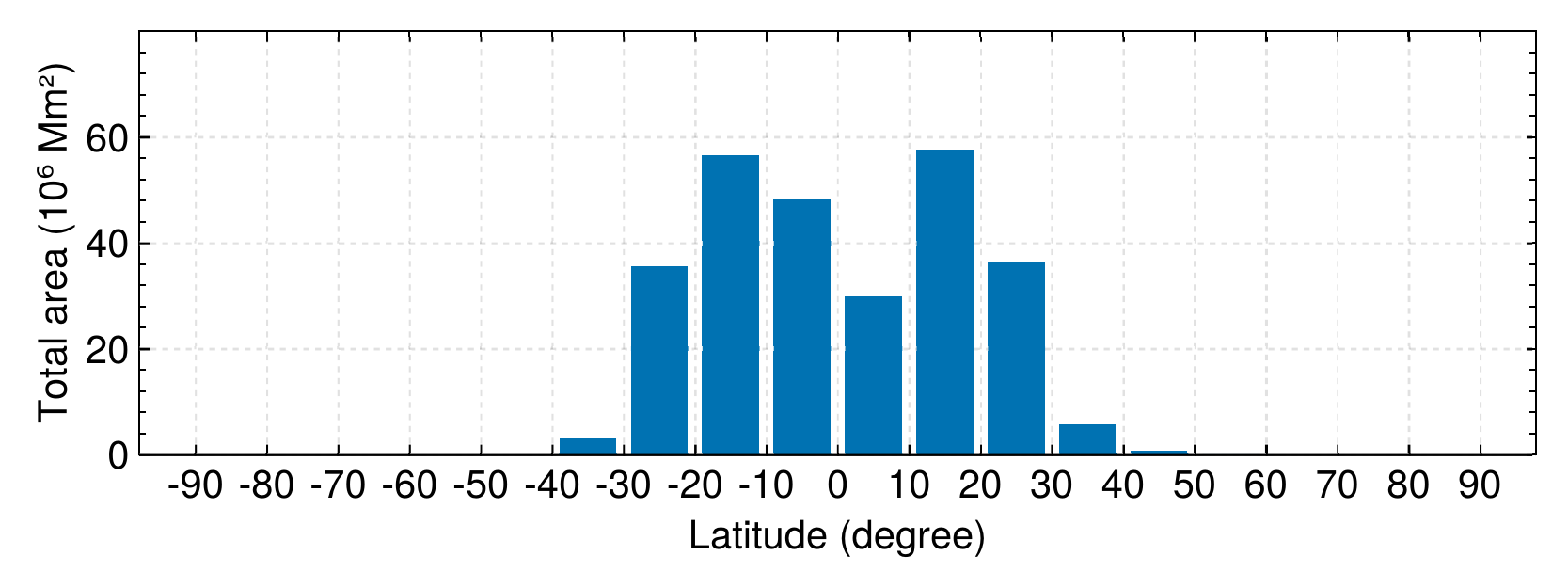}
		\caption{Similar to Figure~\ref{ar_average_area_width_latitude} but for the cumulative area. }
		\label{ar_total_area_latitude}
	\end{figure}

	\begin{figure}[ht!]
		\centering
		\includegraphics[width=\linewidth]{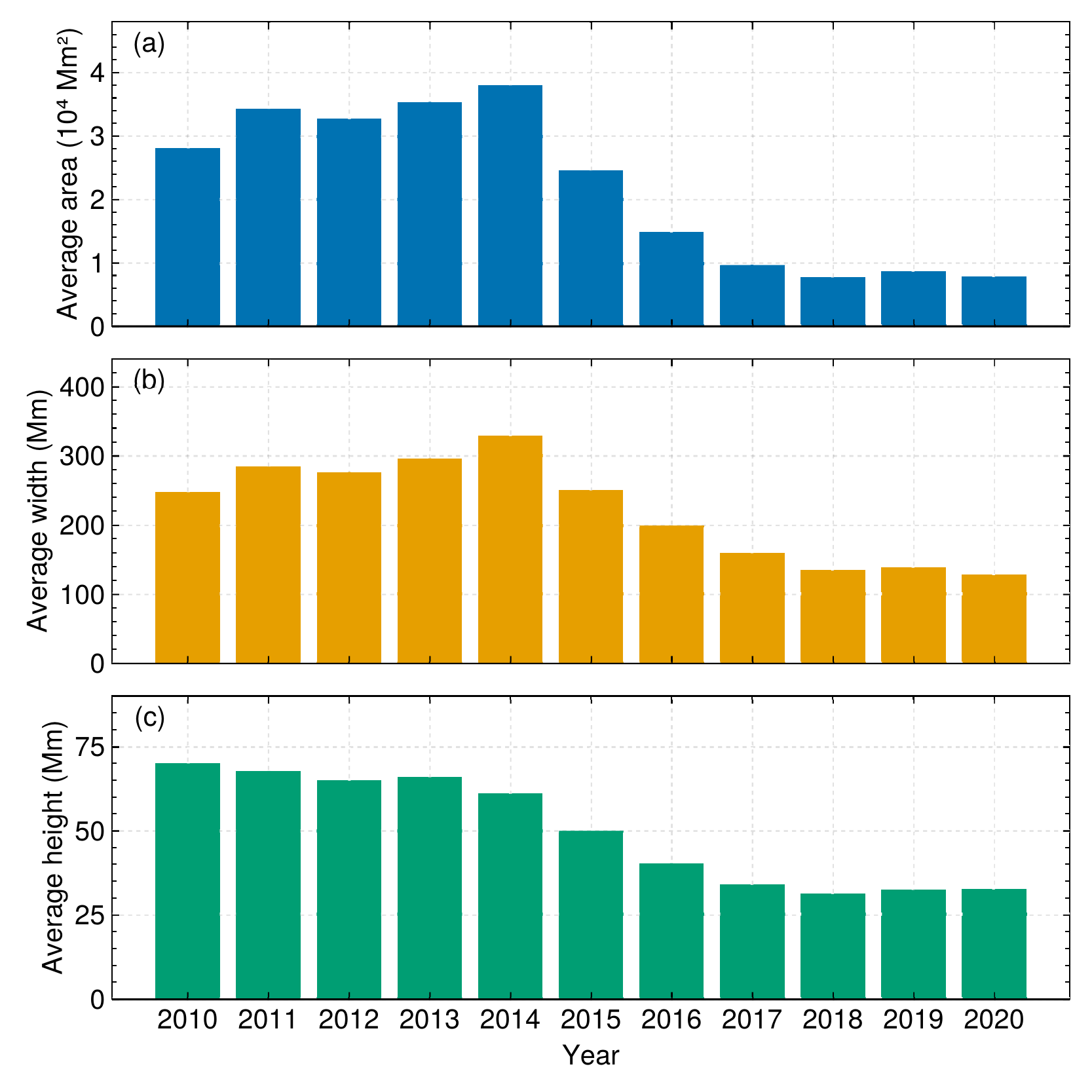}
		\caption{Distributions of the AR features with respect to years. Panels (a), (b), and (c) show the average area, width, and height of the ARs in different years, respectively.}
		\label{ar_average_area_width_height_year}
	\end{figure}

	\begin{figure}[ht!]
		\centering
		\includegraphics[width=\linewidth]{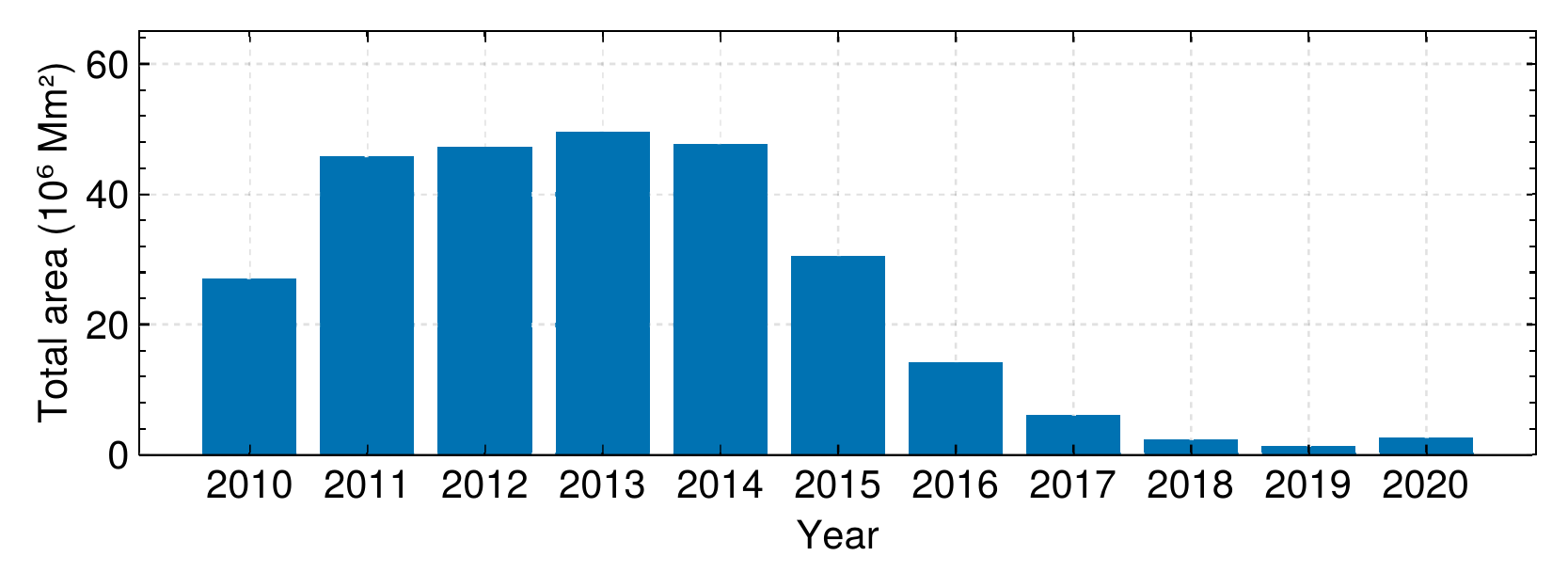}
		\caption{Similar to Figure~\ref{ar_average_area_width_height_year} but for the cumulative area. }
		\label{ar_total_area_year}
	\end{figure}

	As done for prominences, we also analyze the area, height, and width variations of ARs with respect to latitude and time. The definitions of the height and width of ARs are the same as those of prominences. The average areas and widths show bimodal distributions with the highest peaks in the southern hemisphere as shown in Figure~\ref{ar_average_area_width_latitude}. The largest average area is about $4.4 \times 10^{4}~\text{Mm}^2$ and the largest average width is $361~\text{Mm}$, which appears in the latitude band of $0^{\circ}$ to $10^{\circ}$ in the southern hemisphere. The largest average area and average width in the northern hemisphere are smaller than those in the southern hemisphere, which are $3.2 \times 10^{4}~\text{Mm}^2$ and  $268~\text{Mm}$, respectively, and appear in the latitude band of $20^{\circ}$ to $30^{\circ}$. The average height is within the range of 40--60~$\text{Mm}$. The ARs around the equator within the latitude band of $0^{\circ}$ to $10^{\circ}$ have lower heights. The cumulative area of the ARs in different latitude bands also presents a bimodal distribution. The largest cumulative area appears in the latitude band of $10^{\circ}$ to $20^{\circ}$ in both hemispheres. The cumulative area of ARs with latitudes higher than $30^{\circ}$ shows a steep drop in each hemisphere. It indicates that the ARs are mainly formed in the low latitude band of $0^{\circ}$ to $30^{\circ}$ in both hemispheres.

	Figure~\ref{ar_average_area_width_height_year} shows the yearly distributions of the average area, width, and height in panels (a--c). The average area and width both show an increasing trend in the rising phase of solar cycle 24 in general and become a decreasing trend after 2014. There are two peaks of the average area and width, which appear in 2011 and 2014, respectively. The first peaks of the average area and width are $3.4 \times 10^{4}~\text{Mm}^2$ and $281~\text{Mm}$, respectively. The second peaks are $3.8 \times 10^{4}~\text{Mm}^2$ and $324~\text{Mm}$, respectively, which are both larger than the first peaks. However, the average height shows a decreasing trend from the beginning to the end of solar cycle 24 in general. It indicates that the emissions of ARs become gradually weak during solar cycle 24. The evolution of the cumulative area of ARs is plotted in Figure~\ref{ar_total_area_year}. It also shows an increasing trend which is followed by a decreasing trend in solar cycle 24. The largest cumulative area appears in 2013, which is one year later compared to the results of AR number in Figure~\ref{ar_number_year}(a).

\begin{figure}[ht!]
		\centering
		\includegraphics[width=\linewidth]{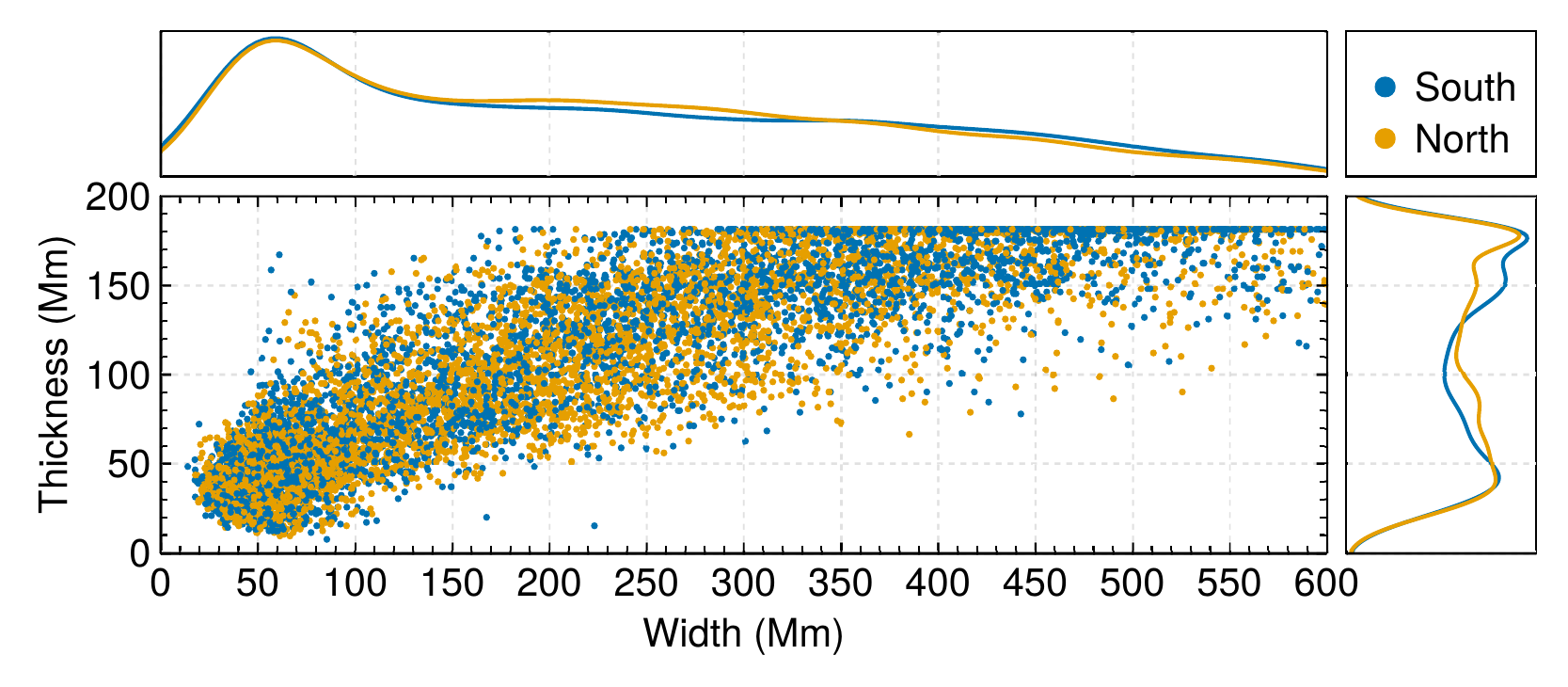}
		\caption{Relationship between the thickness and width of ARs. Each dot represents a single AR. The blue and orange dots represent the observations in the southern and northern hemispheres, respectively. The right panel is the percentage profile with respect to the thickness of the ARs. The top panel is the percentage profile with respect to the width of the ARs. The blue and orange lines represent the results in the southern and northern hemispheres, respectively.}
		\label{ar_height_width}
	\end{figure}

	The relationship between the thickness and width of ARs is shown in Figure~\ref{ar_height_width}. Most of the ARs have widths below $450~\text{Mm}$ and their thickness are below $180~\text{Mm}$. The maximum thickness is due to the limited field of view of the SDO/AIA observations. The right and top panels display the distributions of the ARs thickness and width in the northern (blue line) and southern (orange line) hemispheres, respectively. It is found that the distributions of the thickness and width for ARs are more dispersed than those of prominences. The distributions of the AR width in the northern and southern hemispheres are highly similar. The width distribution peaks at around $55~\text{Mm}$ and the thickness distribution peaks at around $37~\text{Mm}$ and $182~\text{Mm}$. As mentioned above, the largest peak at $182~\text{Mm}$ is due to the limit of the SDO/AIA field of view, which indicates that these ARs have a thickness larger than $182~\text{Mm}$.

\begin{figure}[ht!]
		\centering
		\includegraphics[width=\linewidth]{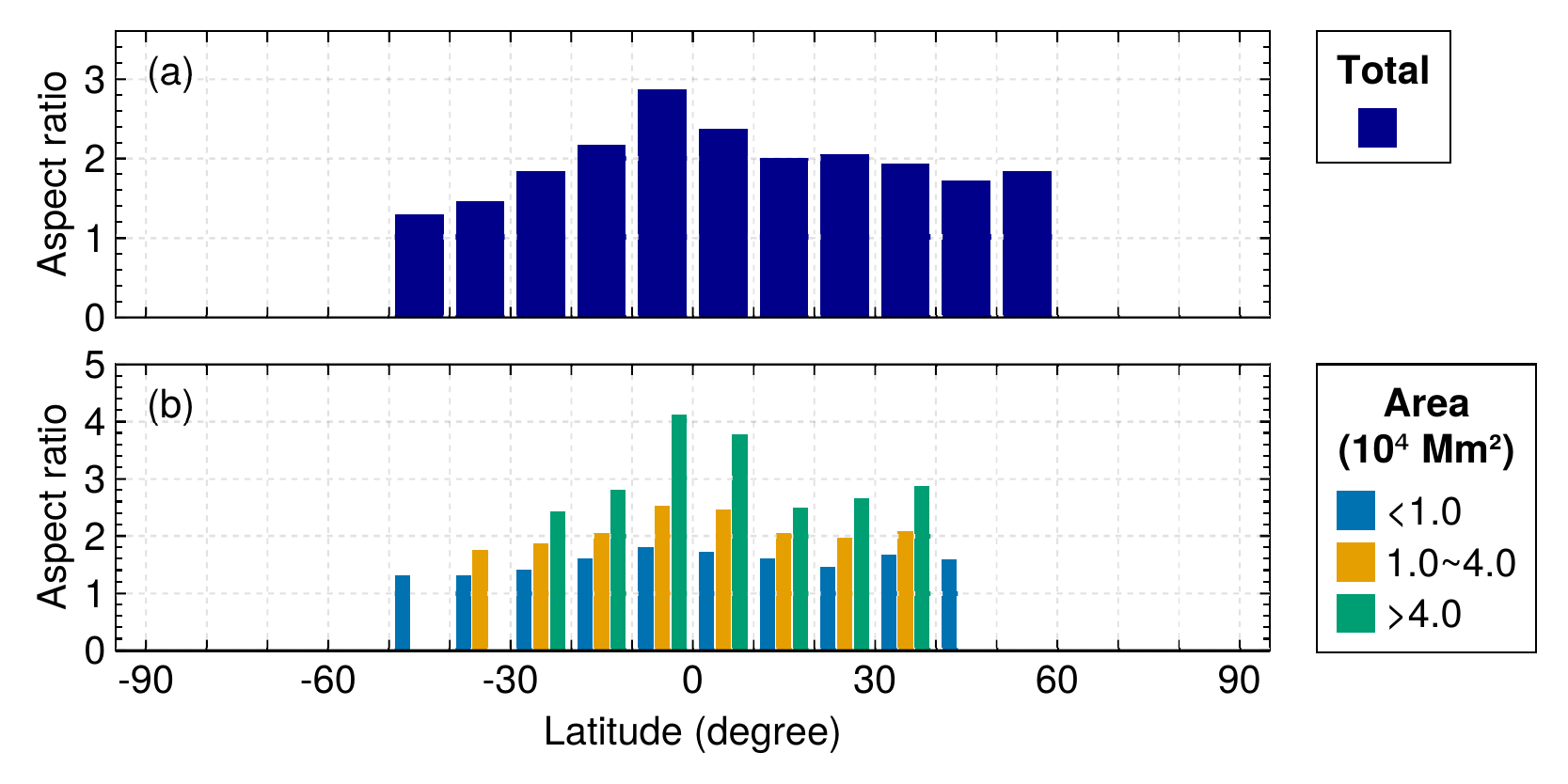}
		\caption{Variation of the aspect ratio of the ARs with respect to latitude.}
		\label{ar_wh_ratio_latitude}
	\end{figure}

\begin{figure}[ht!]
		\centering
		\includegraphics[width=\linewidth]{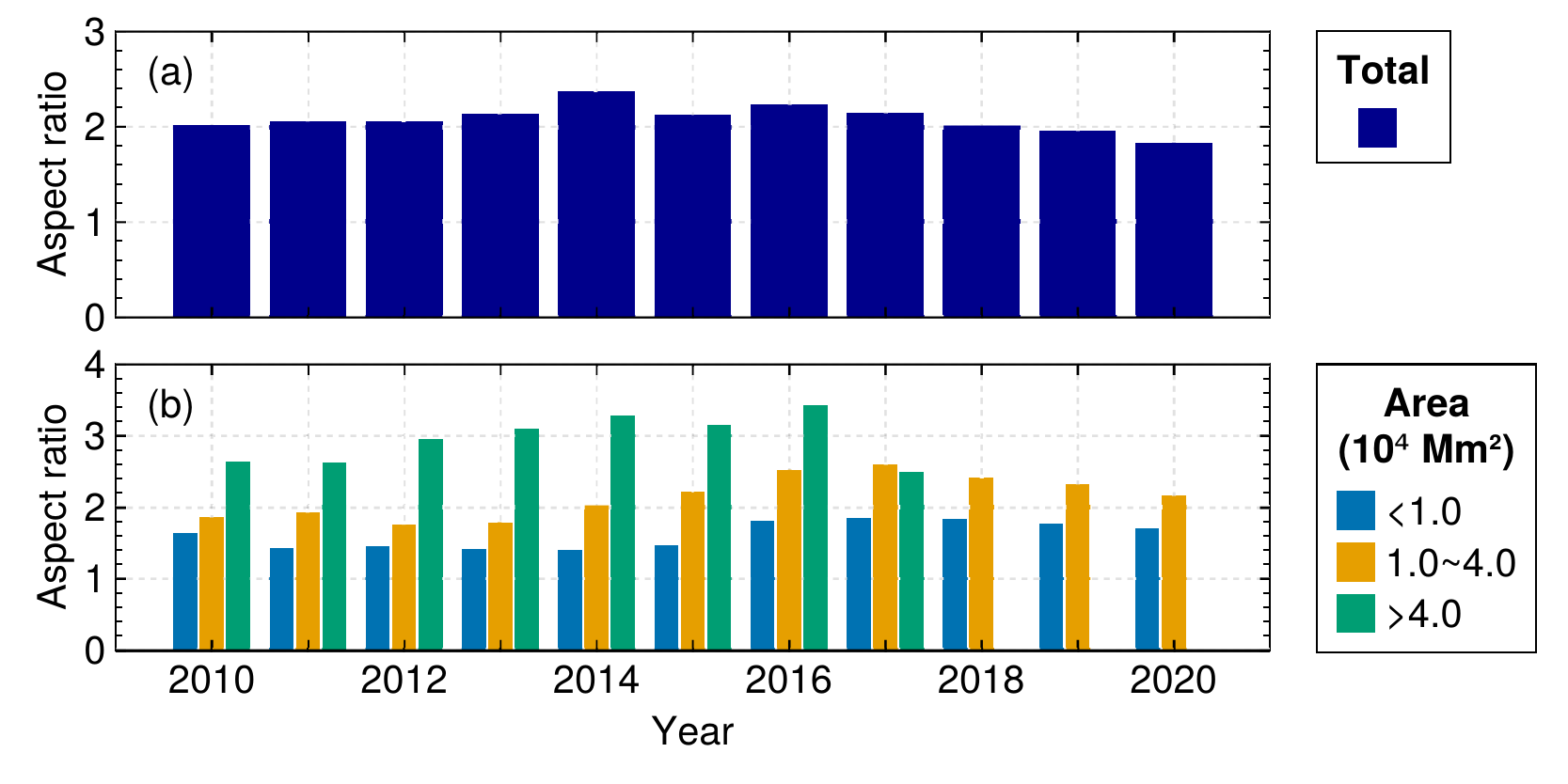}
		\caption{Variation of the aspect ratio of the ARs with respect to years.}
		\label{ar_wh_ratio_year}
	\end{figure}

	Figures~\ref{ar_wh_ratio_latitude} and \ref{ar_wh_ratio_year} show the variations of the aspect ratio of the ARs in different latitude bands and time. The aspect ratio appears to decrease with the increasing latitudes in the southern hemisphere. Unlike the southern hemisphere, the aspect ratio does not show any significant variations with the latitude in the northern hemisphere. The ARs in low latitude bands near the equator have larger aspect ratios. It indicates that the ARs in low latitude bands are wider or thinner. The ARs with areas lower than $1 \times 10^{4}~\text{Mm}^2$ do not show significant variations in the aspect ratio. The ARs with areas larger than $1 \times 10^{4}~\text{Mm}^2$ show a similar variation trend to that of the total ARs. Figure \ref{ar_wh_ratio_year}(a) shows that the aspect ratio of ARs is quite stable at around 2 throughout solar cycle 24, with only two insignificant peaks in 2014 and 2016 and a decreasing trend after 2016. The aspect ratio of the ARs with areas lower than $1 \times 10^{4}~\text{Mm}^2$ is stable during the rising phase of solar cycle 24, then increases after 2015 and remains stable again in the declining phase. The aspect ratio of the ARs with areas within $1 \times 10^{4}$--$4 \times 10^{4}\ \text{Mm}^2$ has a trend of increasing first and decreasing later, with peaks in 2011 and 2017. However, the aspect ratio of the ARs with areas larger than $4 \times 10^{4}~\text{Mm}^2$ increases from 2010 to 2016 in general. The absence of their aspect ratio after 2017 is due to the scarcity of the larger ARs with areas over $4 \times 10^{4}\ \text{Mm}^2$ in these years.

\subsubsection{N-S Asymmetry of Active Regions}

	\begin{figure}[ht!]
		\centering
		\includegraphics[width=\linewidth]{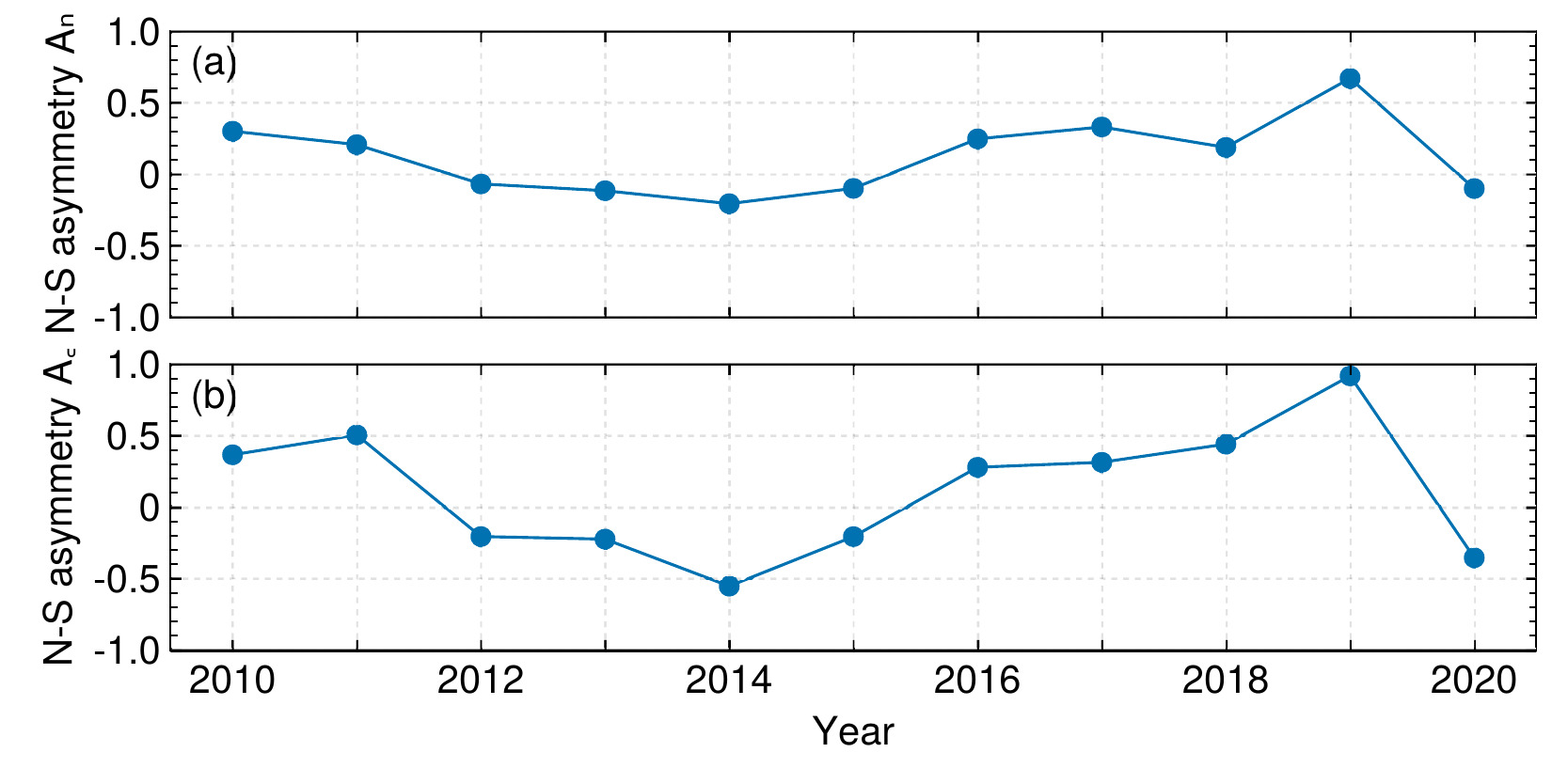}
		\caption{Yearly asymmetries of the number (panel a) and cumulative area (panel b) of all ARs.}
		\label{ar_nsa}
	\end{figure}

	The yearly asymmetries of the AR number and the normalized cumulative area in different latitude bands are plotted in Figure~\ref{ar_nsa}. The asymmetry indices of the AR number and cumulative area indicate that the northern hemisphere is dominant in terms of all ARs, except for 2012--2015,  and 2020. The number of ARs changes with time more drastically than that of prominences. The asymmetry index of the cumulative AR area has a similar trend to that of AR number but with notable asymmetry indices in the years 2011, 2014, and 2019.

\section{Discussion and Conclusions}\label{discussion&conclusion}

	Solar prominences and active regions are important phenomena, their numbers, areas, and migration are the indicators of the changing global magnetic distribution on the Sun \citep{Hathaway2010, Shimojo2013, Gopalswamy2016}. The automated detection methods provide a fast and efficient way to analyze, track, and even predict the properties of magnetic activities. In this paper, we proposed a new efficient and robust automated detection method for prominences and active regions on the solar limb based on CNNs and then made statistical analyses on their morphological features and long-term variations in solar cycle 24. There are $50,456$ prominences and $10,018$ active regions detected in our study.

	The latitudinal distributions of prominences and AR numbers are bimodal with peaks in the latitude bands of $[20^{\circ}$, $30^{\circ}]$ and $[10^{\circ}$, $20^{\circ}]$ in solar cycle 24 as shown in Figures~\ref{p_number_latitude} and \ref{ar_number_latitude}, respectively. \citet{Hao2015} analyzed the filaments detected in the full-disk H$\alpha$ images during solar cycles 22 and 23, as well as the rising phase of solar cycle 24. They found that the latitudinal distribution of the filament numbers is bimodal with peaks appearing in the latitude band of $[10^{\circ}$, $30^{\circ}]$ in both hemispheres throughout the 2.5 solar cycles. \citet{Shimojo2006} studied the automatically detected prominences from the Nobeyama Radiograph data obtained during 1992--2004 and obtained similar results in solar cycle 23. The distribution of the prominences with areas over $4000\ \text{Mm}^2$ is consistent with the total number distribution. The distribution of the prominences with areas in the range 1000--4000 $\text{Mm}^2$ is also bimodal but with peaks in a higher latitude band of $[40^{\circ}$, $50^{\circ}]$. However, the prominences with areas lower than $1000\ \text{Mm}^2$ have more than one peak in higher latitudes in the northern hemisphere. This indicates that the formation of the smallest prominences is more complex in latitudinal behaviors than the larger ones. About $75.9\%$ of prominences are low latitude ones. Similar results were reported by \citet{Wang2010}, \citet{Loboda2015}, and \citet{Hao2015}. The yearly distributions show that the total number of ARs peaks in the year 2011 as shown in Figure~\ref{ar_number_year}(a), and that of prominences peaks in the year 2013, as indicated in Figure~\ref{p_number_year}(a), which is two years later than active regions. This might be related to the fact that the CME occurrence rate lags behind sunspot number by about 1 year \citep{chen11}.

The yearly distribution peaks of the prominences with different areas are synchronous, but those of the active regions are not. The sunspot number shows that there are two peaks in solar cycle 24: the first one is in late 2011 due to the predominance of sunspot emergence in the northern hemisphere and the second peak in around 2014 due to the predominance in the southern hemisphere \citep{Chowdhury2019, Veronig2021, Nandy2021, Du2022, Andreeva2023}. Our results show a similar trend: the limb active regions in the northern hemisphere peak in 2011 and those in the southern hemisphere peak in 2013, as shown in Figure~\ref{ar_number_year}(c). The second peak dominated by the southern hemisphere is one year earlier than that of the sunspot number. This may be due to the detection errors since the AR numbers in 2013 and 2014 are almost the same as shown in Figure~\ref{ar_number_year}(c). The results for the prominences lower than $50^{\circ}$ are similar to that of ARs, but the number difference is not significant between the two hemispheres (Figure~\ref{p_number_year}(c)). However, the high-latitude prominences show that the northern hemisphere is the predominant one from 2010 to 2015.

	The latitude migrations of prominences and active regions are displayed in Figures~\ref{butterfly_and_driftV_total}--\ref{butterfly_and_drift_low_big} and \ref{butterfly_and_driftV_ar}, respectively. While the active regions present a butterfly diagram clearly, the buttery diagram of the prominences is much more scattered. The butterfly diagram of the prominences is consistent with those of other authors \citep{Shimojo2013, Hao2015, Gopalswamy2016, Chatterjee2017}, i.e., two wings in high latitudes rush to the poles and another two wings in low latitudes migrate toward the equator. \citet{Gopalswamy2016} plotted prominence eruption (PE) locations detected by their automated detection method \citep{Gopalswamy2015}, which were also from the 304 \AA \ limb data obtained by SDO/AIA from 2010 to 2016. They found that there appear multiple rush-to-the-pole surges in the northern hemisphere and a steady chain of PEs in the southern hemisphere. Our results shown in Figure~\ref{butterfly_and_driftV_total}(a) are consistent with their findings. Besides, they found that the PEs with latitude higher than $60^{\circ}$ start in 2010 (2012) and end in later 2015 (2014) in the northern (southern) hemisphere. Our results are in agreement with them in the northern hemisphere but not in the southern hemisphere. It is noted that our detection results include both erupting prominences and quasi-static prominences.

The butterfly diagram of the ARs shows two clear wings which are shaded with two blue belts in the latitudes below  $40^{\circ}$ as shown in Figure~\ref{butterfly_and_driftV_ar}(a). In addition, we can see a few detected active regions in the higher latitudes. In fact, these ARs with latitude higher than  $50^{\circ}$ account for less than $1\%$ of all detected ARs, they are probably decayed active regions.

	We took the monthly average latitudes of the prominences and ARs in both hemispheres to analyze the latitudinal migration and drift velocities quantitatively. Cubic polynomial functions were used to fit the drift velocities of the prominences and ARs. We also introduced an approach to ensure that only the data around the fitting curve are taken into account. The results for the high-latitude prominences are displayed in Figures~\ref{butterfly_and_driftV_total}(b--e). The high-latitude prominences mainly migrate toward the polar region with higher velocities in the rising phase of solar cycle 24 in both hemispheres. The drift velocities accelerate from the year 2010 to around later 2013 in both hemispheres, and then decrease later in the southern hemisphere. The velocities range from $0.5~\text{m}\ \text{s}^{-1}$ (or $0.003^{\circ}~\text{day}^{-1}$) to $8.4~\text{m}\ \text{s}^{-1}$ (or $0.06^{\circ}~\text{day}^{-1}$). The average velocities are $4.2~\text{m}\ \text{s}^{-1}$ (or $0.03^{\circ}~\text{day}^{-1}$) and $2.8~\text{m}\ \text{s}^{-1}$ (or $0.02^{\circ}~\text{day}^{-1}$) in the northern and southern hemispheres, respectively, which are on the order of magnitude of the photospheric meridional flows.

\citet{Xu2018} manually identified polar crown filaments from 1973 to 2018 covering solar cycles 21 to 24 from the H$\alpha$ data recorded by Kanzelh\"{o}he Solar Observatory and Big Bear Solar Observatory. Their analysis showed a clear poleward migration pattern of polar crown filaments and the drift velocities are about $0.4^{\circ}$ to $0.7^{\circ}$ per Carrington rotation. In their results, the migrate velocities of polar crown filaments in solar cycle 24 are $0.55^{\circ}\pm0.10^{\circ}/\text{rotation}$ and $0.63^{\circ}\pm0.08^{\circ}/\text{rotation}$ in the northern and southern hemispheres, respectively. \citet{Diercke2019} adopted an automated method to detect polar crown filaments from 2012 to 2018 from the data observed by the Chromospheric Telescope at the Observatorio del Teide. The derived migration velocity is $0.79^{\circ}\pm0.08^{\circ}/\text{rotation}$ in the southern hemisphere in solar cycle 24, which is slightly larger than that of \cite{Xu2018} due to the incomplete data. Our results are in general agreement with \cite{Xu2018}, but the drift velocity in the northern hemisphere is higher than that in the southern hemisphere in our results. \citet{Tlatova2020} analyzed the drift of the polar prominences in solar cycles 23 to 24 detected by image processing technique from the observations obtained by various observatories. They found that the drift velocity of prominences is in the range of 2--7 $\text{m}\ \text{s}^{-1}$. The average drift velocity of the prominences in the latitude band of $60^{\circ}$ to $80^{\circ}$ is $7.07~\text{m}\ \text{s}^{-1}$ and $6.06~\text{m}\ \text{s}^{-1}$ in the northern and southern hemispheres during solar cycle 24, respectively. The difference in the velocities may be due to the data sets with different latitude bands and time ranges.

	The distribution of low-latitude prominences is relatively more scattered as shown in Figure~\ref{butterfly_and_driftV_total}(a). We can not obtain a converging fitting range with the same approach used for high-latitude prominences, so we fit all the low-latitude prominences directly. The low-latitude prominence migrates toward the equator with velocities no more than $0.7~\text{m}\ \text{s}^{-1}$ (or $0.005^{\circ}~\text{day}^{-1}$) in the rising phase of solar cycle 24. After around 2015, they migrate toward high latitudes and the monthly average latitudes of prominences become relatively divergent as shown in Figures~\ref{butterfly_and_driftV_total}(b) and (d). The drift velocities are relatively low and steady in the northern hemisphere, while they accelerate in the southern hemisphere though the velocity is still no more than $1.4~\text{m}\ \text{s}^{-1}$.

In addition, we further divided the prominences into three groups according to their areas to analyze their drift velocities in more detail.  It is revealed that the larger the areas of the prominences are, the clearer the wings of the butterfly diagram are. Despite the prominences with areas lower than  $1000\ \text{Mm}^2$ having a much wider latitudinal distribution, we still calculated their drift velocities. The high-latitude prominences with areas lower than $4000\ \text{Mm}^2$  migrate toward the polar regions and the drift velocities show drastic variations in the northern hemisphere which can be over $10~\text{m}\ \text{s}^{-1}$. On the contrary, the high-latitude prominences with areas lower than $4000\ \text{Mm}^2$ in the southern hemisphere migrate toward the polar region with a low velocity in the rising phase of solar cycle 24, which is accelerated to $5.6~\text{m}\ \text{s}^{-1}$ around the maximum of solar cycle 24. The high-latitude prominences with areas larger than $4000\ \text{Mm}^2$ also migrate toward the polar regions. Their drift velocities increase in the rising phase and reach a maximum of around $5.6~\text{m}\ \text{s}^{-1}$ around the maximum of solar cycle 24 in later 2013. Due to the difference in the fitting time intervals, the drift velocity of the prominences in the southern hemisphere shows a deceleration toward the polar region afterward.

The low-latitude prominences with different area ranges have similar trends: they first migrate toward the equator with a decreasing velocity, and then migrate toward higher latitudes after the solar maximum with an increasing velocity. The drift velocities of the low-latitude prominences with areas lower than $4000\ \text{Mm}^2$ show obvious asymmetry between two hemispheres while those of the prominences with areas larger than $4000\ \text{Mm}^2$ have similar trends in both hemispheres. Besides, the derived drift velocities of the prominences with areas lower than $1000\ \text{Mm}^2$ are lower than $0.56~\text{m}\ \text{s}^{-1}$, but that for the prominences with areas in the range of 1000--4000 $\text{Mm}^2$ are higher and can be up to $0.8~\text{m}\ \text{s}^{-1}$.  The large prominences with areas larger than $4000\ \text{Mm}^2$ also have higher drift velocities, especially for those in the southern hemisphere, which are up to $2.0~\text{m}\ \text{s}^{-1}$. This indicates that the migration is much more significant for larger prominences.

	\citet{Li2010} analyzed the Carte Synoptique solar filaments archive observed by Meudon Observatory from 1919 to 1989 corresponding to solar cycles 16 to 21 and found that the latitudinal drift velocity of the filaments ranges within about $1.5$--$3.5~\text{m}\ \text{s}^{-1}$, and there is a slight difference between two hemispheres within a cycle. A similar velocity range during solar cycle 22 and 23 was reported by \cite{Hao2015}.  \citet{Li2010} also found that the latitudinal drift velocity of the filaments first decreases and then increases within a cycle, where the decreasing time is about $7.5$ years and the increasing time is about $3.5$ years. Our results show a similar trend but the maximum drift velocity is slightly lower than theirs and there is obvious asymmetry between the northern and southern hemispheres.

	ARs migrate toward the equator throughout the whole solar cycle 24 with a relatively low velocity within the range of $0.5$--$2.8~\text{m}\ \text{s}^{-1}$, which is quite consistent with the derived velocity of sunspots in \citet{Li2010}. They also showed deceleration in the rising phase, acceleration in the declining phase, and north-south asymmetry. In the rising phase, the drift velocity in the northern hemisphere is higher than that in the southern hemisphere from 2010 to 2014. After 2015, the velocity increases in both hemispheres, but the migration in the southern hemisphere is faster than that in the northern hemisphere. The asymmetry of the drift velocity in the northern and southern hemispheres for the low-latitude prominence and ARs may be related to the unusual migration of the magnetic field in solar cycle 24 which has been reported by many authors \citep{Shimojo2013, Sun2015, Gopalswamy2016, Janardhan2018}.

	Both ARs and prominences are 3D structures. However, what we observe are the 3D structures projected on the plane of the sky. In other words, the derived features suffer from the projection effects. \citet{Hao2015} found that about $80\%$ of filaments have tilt angles (the angle between the filament spine orientation and its local latitude) within $[0^{\circ}, 60^{\circ}]$. Furthermore, \citet{Tlatov2016} revealed that the tilt angles of polar crown filaments are near zero or negative (note that being negative means that the filament spine is tilted toward the equator rather than the poles). Therefore, frequently we see one footpoint of a prominence on the solar disk and the other behind the limb. In this sense, the length of a prominence is meaningless due to the projection effects except when the prominence is along the meridian direction. In our study, we chose the area, width (the span of the prominence along the solar limb), and height (the geometric center height in the radial direction) as the characteristic features of prominences. Besides, it should be noted that we processed one image per day, which means that the features of the prominences change day by day during the period when they are detected even if a 3D prominence does not change in reality. Nevertheless, our results are still statistically significant.

	The yearly and latitudinal average area, width, and height of prominences are within the ranges of 1800--3600 $\text{Mm}^2$, 70--130 Mm, and 15--23 Mm, respectively. \citet{Wang2010} applied their SLIPCAT method to detect the prominences from the STEREO/EUVI 304  \AA \ data during 2007--2009. They reported that their well-tracked prominences have an average area of  $2681\ \text{Mm}^2$ and an average centroid height of $25.55 \ \text{Mm}$. Their area is consistent with ours but our derived average height is smaller than theirs. The difference results from the threshold for removing small regions: the threshold is $500\ \text{Mm}^2$ in their study whereas ours is $250\ \text{Mm}^2$. It implies that our results would contain smaller prominences which have lower heights. It is also found that the height distribution peaks around $25 \ \text{Mm}$. Similar width and height of prominences were also derived by \citet{Loboda2015}, who also used an automated detection method to the quiescent and quiescent-eruptive prominences observed by TESIS/FET \citep{Kuzin2009} during 2008--2009. In their analysis, eruptive prominences have larger heights than those of quiescent prominences. The latitudinal distributions of the prominence average area, width, and cumulative area are bimodal with the peaks appearing in the latitude band of $[20^{\circ}, 30^{\circ}]$.  These results are consistent with those of solar filaments \citep{Hao2015} since they are the same features viewed from different perspectives. The bimodal latitudinal distribution of the prominence area suggests that the prominences in low-latitude bands have larger projected areas than those in high latitudes in both hemispheres. It indicates that the high-latitude prominences have smaller tilt angles than those in low-latitude bands, which conforms to the result of \citet{Tlatov2016}. In fact, the latitudinal distribution of the prominence width has the same bimodal distribution, which also supports this deduction, since the width represents the angular span of a prominence along the solar limb.

	The cumulative area decreases more rapidly with the increasing latitude than that of the average area.  However, the average height gradually increases with the increasing latitude and has peaks around the polar regions in the high latitude bands. It indicates that the prominences in high latitudes are higher. The average area and width show decreasing trends from the beginning to the end of solar cycle 24 in general, although the differences are not significant from 2011 to 2015.  The maximum average area and width are $3408~\text{Mm}^2$ and  $120~\text{Mm}$, respectively, which appear in the years 2011 and 2013, respectively. The average height gradually increases from the year 2010 to 2014, which peaks at $21~\text{Mm}$ in 2014. After that, it decreases in the following two years and maintains a stable value around $15~\text{Mm}$. It indicates that prominences become lower in the declining phase of solar cycle 24. The largest cumulative area appears in 2013, which is consistent with the result of the prominence number.

	We also calculated the distance between the bottom-height of the detected prominence and the solar limb. There are only $10\%$ prominences completely suspended in the corona totally detached from the solar limb. This result is derived from the 304 \AA\ observations, and whether other prominences are really attached to the solar limb requires comparisons with other observations, such as the typical H$\alpha$ images. Since prominences and filaments are more extended in EUV than that in  H$\alpha$ \citep{Heinzel2001, Wang2010}, it can be inferred that prominences are much more visible in 304\ \AA \ than that in H$\alpha$, some EUV prominences may be invisible in H$\alpha$. Besides, some really detached prominences might be identified as attached to the solar limb due to the projection effects.

	To our knowledge, there are not many studies on the automated detection of ARs above the solar limb. In this paper, we derived the morphological features of ARs in the same way as for prominences. The yearly and latitudinal average area,  width, and height of ARs are within the ranges of $2.3 \times 10^3$--$4.4\times 10^4 \ \text{Mm}^2$, 60--370 Mm, and 30--70 Mm, respectively. The average area, width, and cumulative area show overall bimodal distributions with respect to latitude. The ARs around the equator within the latitude band of $[0^{\circ}, 10^{\circ}]$ have lower heights. In fact, $97\%$ of the detected ARs have latitudes lower than $40^{\circ}$. As for the latitudinal distribution of the cumulative area, over $95\%$ of the cumulative AR areas are typically in the latitudes lower than $40^{\circ}$ in both hemispheres.  It is consistent with the distribution of the ARs represented by the sunspot groups on the solar disk \citep{Li2010}. The average area and width show an increasing trend in the rising phase of solar cycle 24 in general, which becomes a decreasing trend after the solar maximum. Both the average area and width have two peaks which appear in the years 2011 and 2014, respectively. The cumulative area shows a similar trend but has only one peak in 2013, which is one year later compared to that of the AR number. The average height gradually decreases with time throughout solar cycle 24.

	The N-S asymmetry indices of the prominence number, the cumulative prominence area, the AR number, and the cumulative AR area indicate that the northern hemisphere is the dominant hemisphere in solar cycle 24, which is in accord with the results for sunspots, magnetic morphological classifications of ARs and filaments \citep{Hao2015, Chowdhury2019, Veronig2021, Chandra2022, Zhukova2023}. The indices for the low-latitude prominences have the same trend. On the other hand, while the high-latitude prominences in the rising phase from 2010 to 2015 show that the dominant hemisphere is the northern one, it changes to the southern hemisphere after 2015 during the declining phase of solar cycle 24. The asymmetry indices of the AR number and the cumulative area are larger than that of the prominences, indicating that the asymmetry is stronger for ARs. The yearly variations of the AR number in the two hemispheres (Figure~\ref{ar_number_year}(c)) are not synchronous: while the northern hemisphere peaks in 2011, the southern hemisphere peaks in 2013. Correspondingly, the northern hemisphere dominates in the years 2010 to 2011 and the southern hemisphere dominates in the following 2012 to 2015. Such a result is consistent with the N-S asymmetry indices of the sunspot number and the cumulative sunspot number from the rising phase to the maximum of solar cycle 24 \citep{Chowdhury2019}, which indicates that the detected ARs above the solar limb can also serve as an index characterizing the strength of solar cycles.

	In the U-Net model, the calculated parameters depend on the size of the input images, kernel sizes, network depth, and other parameters. A larger model apparently requires a larger GPU memory. Due to the limitation of our hardware, during pre-processing, we have to resize the image size to 192 by 4608, about a quarter in area of their original size. We will utilize more computational resources to improve our model. Besides, our model is a supervised one, which means its performance depends on the provided data set. We can improve the performance by providing a larger and more precise data set until we reach the margin. In addition, our pipeline can be extended to process data from other telescopes such as CHASE \citep{Li2022, qiu22} and ASO-S \citep{Gan2019} in prominence and AR detection and tracking.

In summary, we introduced an efficient and accurate method based on deep learning to detect prominences and ARs in 304 \AA\ images observed by SDO/AIA during solar cycle 24. Most of our statistical results based on the new method are in agreement with previous studies, which guarantees the validity of our method.  The key results are listed as follows.

(1) The maximum prominence number and the maximum cumulative prominence area occur in the year 2013.  The prominences with latitudes lower than $50^{\circ}$ occupy $76\%$ of all prominences and the occurrence of these prominences peaks in 2011 (2013) in the northern (southern) hemisphere. The latitudinal distribution of the prominence number is bimodal, with two peaks in the latitude band of $[20^{\circ}, 30^{\circ}]$ in each hemisphere.

(2) The latitudinal distributions of prominence average area, width, and cumulative area are also bimodal and the peaks are at the latitude band of $[20^{\circ}, 40^{\circ}]$ in each hemisphere. The average area, width, and height remain stable during the rising phase and decrease from the year around the maximum to the end of solar cycle 24.

(3) Most prominences in the latitudes higher than $50^{\circ}$ migrate toward the polar regions most of the time, especially in the rising phase of solar cycle 24, with a velocity from $0.5~\text{m}\ \text{s}^{-1}$ (or $0.003^{\circ}~\text{day}^{-1}$) to $11.2~\text{m}\ \text{s}^{-1}$ (or $0.08^{\circ}~\text{day}^{-1}$).  It was found that the velocity increases in the rising phase and starts to decrease after the solar maximum of solar cycle 24. There are significant N-S asymmetries in the drift velocities of the high-latitude prominences within different area ranges.

(4) The prominences with latitudes lower than $50^{\circ}$ migrate toward the equator most of the time with a relatively low velocity ranging from $0.3~\text{m}\ \text{s}^{-1}$ (or $0.002^{\circ}~\text{day}^{-1}$) to $1.4~\text{m}\ \text{s}^{-1}$ (or $0.01^{\circ}~\text{day}^{-1}$). The latitudinal migration of the low-latitude prominences has three stages in solar cycle 24: they drift toward the equator with a decreasing velocity in the rising phase of the solar cycle, then remain stationary around the solar maximum, and then drift poleward with an increasing velocity in the declining phase. The larger the areas of the prominences are, the higher the drift velocities are. There is significant N-S asymmetry in the drift velocities between low-latitude prominences with areas lower than $4000\ \text{Mm}^2$.

(5) The N-S asymmetry indices of the total and low-latitude prominence number and the cumulative area indicate that the northern hemisphere is the dominant hemisphere in solar cycle 24, though the asymmetry is quite weak. The high-latitude prominences show much stronger N-S asymmetry, i.e., the northern hemisphere is dominant in $\sim$2011 and $\sim$2015, but the southern hemisphere is dominant during 2016--2019.

(6) The maximum of the total AR number and the maximum cumulative AR area occur in the years 2012 and 2013, respectively.  There are $99\%$ ARs with latitudes lower than $50^{\circ}$ and their latitudinal distribution peaks in the latitude band of $[10^{\circ}, 20^{\circ}]$ in both hemispheres. The peak of AR number appears in 2011 (2013) in the northern (southern) hemisphere.

(7) The latitudinal distributions of the AR total number, average area, width, and cumulative area are almost bimodal and their peaks are in the latitude band of $[10^{\circ}, 30^{\circ}]$ in both hemispheres. The average area and width show an increasing trend in the rising phase of solar cycle 24 in general, and become a decreasing trend after the solar maximum. Both the average area and the width of ARs have two peaks, which appear in the year 2011 and 2014.

(8) ARs migrate toward the equator most of the time with a relatively low velocity from $0.2~\text{m}\ \text{s}^{-1}$ (or $0.001^{\circ}~\text{day}^{-1}$) to $2.8~\text{m}\ \text{s}^{-1}$ (or $0.02^{\circ}~\text{day}^{-1}$). The latitudinal migration of ARs has two phases in solar cycle 24: from the beginning to the year after the solar maximum, the drift velocity shows deceleration, then turns to accelerate in the declining phase. The velocity is larger in the northern hemisphere during the former phase and becomes larger in the southern hemisphere during the latter phase.

(9) The asymmetry indices of the AR number and cumulative area indicate that the northern hemisphere is dominant in terms of all ARs, except for 2012--2015,  and 2020. The number of ARs changes with time more drastically than that of prominences.

\begin{acknowledgements}
	We are grateful to the SDO teams for providing the observational data. We also thank the referee very much for the constructive suggestions which greatly improved the paper in various ways. This work was supported by the National Key Research and Development Program of China (2020YFC2201200) and NSFC under grants 12173019 and 12127901, as well as the AI \& AI for Science Project of Nanjing University.
\end{acknowledgements}

%\bibliography{manuscript}{}
%\bibliographystyle{aasjournal}

\end{document}